\documentclass[11pt]{article}
\usepackage{amsmath, amsthm, amssymb}
\usepackage{geometry}
\usepackage{booktabs}
\usepackage{float}
\usepackage{array}
\usepackage{xcolor}
\usepackage{tikz}
\usetikzlibrary{positioning}
\definecolor{cursedcol}{RGB}{178,24,43}
\definecolor{classicalcol}{RGB}{33,102,172}
\definecolor{cursedfill}{RGB}{238,150,132}
\definecolor{classicalfill}{RGB}{226,238,248}
\definecolor{classicalbar}{RGB}{198,219,239}
\usepackage[hidelinks]{hyperref}
\usepackage[round]{natbib}
\usepackage{setspace}
\theoremstyle{plain}
\newtheorem{theorem}{Theorem}[section]
\newtheorem{proposition}[theorem]{Proposition}
\newtheorem{corollary}[theorem]{Corollary}
\newtheorem{lemma}[theorem]{Lemma}

\theoremstyle{definition}
\newtheorem{definition}[theorem]{Definition}
\newtheorem{example}[theorem]{Example}

\theoremstyle{remark}

\newcommand{\supp}{\operatorname{supp}}
\newcommand{\marg}{\operatorname{marg}}
\newcommand{\pcell}[2]{\begin{array}{@{}c@{\hspace{1.4em}}c@{}}\rule{0pt}{2.4ex} & #2\\ #1 & \rule[-1ex]{0pt}{1ex}\end{array}}

\begin{document}
\renewcommand{\thefootnote}{\fnsymbol{footnote}}
\begin{center}
\textbf{\Large{}Cursed Rationalizability}\\[12pt]
{\large Shani Cohen\footnote{The Hebrew University of Jerusalem. Email: s.cohen@mail.huji.ac.il} \qquad Shengwu Li\footnote{Harvard University. Email: shengwu\_li@fas.harvard.edu}}\\[8pt]
September 23, 2026
\par\end{center}
\renewcommand{\thefootnote}{\arabic{footnote}}
\setcounter{footnote}{0}
\begin{abstract}
\noindent Cursed players understand the distribution of opponents' actions and the distribution of the payoff-relevant state, conditional on their own information, but respond as if the two were independent. Cursed equilibrium additionally requires statistically correct beliefs about the distribution of play. We propose a solution concept that separates the bias from equilibrium: cursed rationalizability iteratively deletes every action that is not a cursed response to a conjecture about opponents' surviving play. The cursed-rationalizable set is the interim correlated rationalizable set of a virtual game, in which payoffs are averaged over the state. Under strategic complements, cursed rationalizability yields the same bounds on behavior as cursed equilibrium. A cautious refinement, the analog of admissibility, delivers sharp predictions in the laboratory games that documented the winner's curse. 
\end{abstract}

\section{Introduction}\label{sec:intro}

People often neglect what other people's behavior reveals about what other people know. Consider a common-value auction: You win the object only if your bid exceeds every other bid. If so, your estimate of the object's value probably exceeds everyone else's estimate. It follows that your estimate is probably too high. In practice, many bidders fail to account for this, and bids are so high that winning is unprofitable, a phenomenon known as the ``winner's curse". 

The winner's curse is robust in the laboratory. It survives public information, high stakes, and moderate experience, fading only for super-experienced bidders \citep{kagellevin2016,kagellevin1986,kagelrichard2001}. Nor is the mistake confined to auctions: in the takeover game, subjects neglect what a counterparty's willingness to trade reveals about the value of the traded asset, bidding far above the equilibrium price \citep{samuelsonbazerman1985, holtsherman1994}.

One leading approach to formalize this is cursed equilibrium \citep{eysterrabin2005}. Cursed equilibrium posits that each player understands the distribution of opponent actions and the distribution of the payoff-relevant state, conditional on his own type. But the opponent actions may be correlated with the state, because the opponents have private information. When computing his own response, a cursed player neglects this correlation, behaving as if opponent actions and the state are conditionally independent.

Cursed equilibrium is a fixed-point theory: each player's conjecture about the distribution of play is correct. That fixed-point assumption and the inference mistake pull in opposite directions. If players are experienced enough to know the distribution of play, why don't they notice their costly mistakes?\footnote{\citet[p.~1632]{eysterrabin2005} saw the tension---``At first blush, this combination may seem unlikely''---and proposed a learning foundation with partial feedback for the fully cursed case. A learning defense is compelling exactly for experienced players, and the interactions in which cursedness does its explanatory work are often the opposite: a company is acquired once; an offshore tract is auctioned once; a litigant settles or goes to trial once.}

This tension also arises in experimental data. Construction firm managers with real-world bidding experience still fell prey to the winner's curse in the laboratory, performing no better than student subjects \citep{dyerkagellevin1989}. Plausibly, the managers relied on heuristics when bidding for contracts; heuristics that they failed to apply in a new context. Thus, \citet{dyerkagellevin1989} concluded that ``the winner's curse is likely to be strongest in the start-up phase of a market, in those markets which experience the greatest turnover of participants and in markets where large numbers of agents come in and out sporadically so as not to acquire any strong learning from experience.'' Similarly, \citet{fudenberg2006}  argued that ``the fact that the amount of `cursedness' typically declines as subjects become more experienced suggests that the curse, while real, is not an equilibrium phenomenon.''

The data present a dilemma. Cursed equilibrium combines two hypotheses: the inference mistake and equilibrium coordination. In initial play, cursedness is robust, but the equilibrium assumption is implausible. For experienced players, the equilibrium assumption is plausible, but the inference mistake is sometimes diminished.

The goal of this paper is to separate cursedness from equilibrium, and thereby to clarify when equilibrium is needed. Data that reject the joint hypothesis may nonetheless be consistent with cursedness alone. And in some settings, cursedness is identified even without the hypothesis of coordination. 

Classical game theory confronted a similar question, for rational players, and resolved it with rationalizability \citep{bernheim1984, pearce1984}: discard equilibrium beliefs, retain best response, and close the model with common knowledge of rationality. For Bayesian games, the classical benchmark is interim correlated rationalizability \citep{dekelfudenbergmorris2007}: Each type starts with all actions available. Iteratively delete, for each type $t_i$, every action that is not a best response to some belief (about the state $\theta$ and opponents' types $t_{-i}$ and actions) that agrees with his posterior on $(\theta,t_{-i})$ and assigns positive probability only to actions that remain for the corresponding type. The limit of this process is the \emph{interim correlated rationalizable set}.

This paper does the same for cursed players. Say that an action is a \emph{cursed response} to a conjecture about opponents' type-contingent play if it maximizes expected utility when the conjecture's action distribution is paired with the type's posterior over the state as if the two were independent. Iteratively delete actions as in the classical case, but substitute cursed response for best response. The limit of this process is the \emph{cursed-rationalizable set}: for each type, the actions that survive. If action $a_i$ is cursed-rationalizable for type $t_i$, then $a_i$ is a cursed response to a conjecture with support on the cursed-rationalizable set. The cursed-rationalizable set is the behavior consistent with cursed response and common knowledge of cursed response. The concept generalizes to partial cursedness, using a definition of $\chi$-cursed response that interpolates between best response at $\chi = 0$ and cursed response at $\chi = 1$.

Cursed equilibria are equivalent to Bayesian Nash equilibria of a virtual game, in which we modify each player's utility function by averaging over the state $\theta$ conditional on his own type $t_i$. As one might hope, the cursed-rationalizable set of the original game is exactly the interim correlated rationalizable set of the virtual game.  The parallel carries foundations with it. Justification at each stage is a linear-programming problem, so the cursed hierarchy is computable in polynomial time and coincides with iterated strict dominance in the virtual game. Moreover, the final set is invariant to deletion order. 

Interim correlated rationalizability and cursed rationalizability can yield disjoint predictions, as in this example: There are two states, good and bad; the good state has probability $\tfrac34$. Player 1 observes the state and chooses whether to participate; participating pays her $1$ in the good state and $-1$ in the bad state, so she participates exactly in the good state, whatever her conjecture. Player 2 does not observe the state and chooses whether to enter. Staying out pays $0$, and the payoff from entering depends on the state and on player 1's choice:
\[
\begin{array}{c|cc}
\text{payoff from entry} & \theta = \text{good } (\tfrac34) & \theta = \text{bad } (\tfrac14) \\
\hline
\text{player 1 out} & -1 & -5 \\
\text{player 1 in} & \phantom{-}1 & \phantom{-}1
\end{array}
\]
The state matters to player 2 only through player 1's absence---and player 1 is absent exactly in the bad state. A classical reasoner accounts for what absence reveals: entering without player 1 happens only in the bad state, so entry is worth $\tfrac34(1) + \tfrac14(-5) = -\tfrac12$, and once player 1's play is pinned down, the interim correlated rationalizable set is $\{\text{out}\}$. A cursed reasoner holds the same conjecture---player 1 participates exactly in the good state, hence with probability $\tfrac34$---but pairs it with her posterior as if independent: she meets participation with probability $\tfrac34$, worth $1$, and absence with probability $\tfrac14$, worth $\tfrac34(-1) + \tfrac14(-5) = -2$ on average over her posterior. Entry is worth $\tfrac34(1) + \tfrac14(-2) = \tfrac14 > 0$, and the cursed-rationalizable set is $\{\text{enter}\}$.

Every cursed equilibrium strategy profile has support in the cursed-rationalizable set, and the containment can be strict. Cursed rationalizability allows for some costly speculative bets that are ruled out by interim correlated rationalizability and \textit{also} ruled out by cursed equilibrium. In essence, cursed equilibrium requires each type's conjecture to be the conditional of the same distribution of play---namely, the equilibrium distribution. Tying conjectures together in this way prevents cursed response from rationalizing some patterns of speculative betting. Section~\ref{sec:vsce} develops the example formally.

While rationalizability can in general be quite permissive, it has sharp predictions in games of strategic complements \citep{milgromroberts1990}. We show that the same is true for cursed rationalizability. In supermodular Bayesian games---on lattice action sets---the cursed-rationalizable set of every type has a largest and a smallest action, and the extremal profiles are themselves fully cursed equilibria. Therefore, under strategic complements, cursed rationalizability yields the same bounds on behavior as cursed equilibrium. 

Level-$k$ models also discard equilibrium, anchoring beliefs on a na\"{i}ve level-0 and iterating best responses finitely many times \citep{nagel1995, stahlwilson1994, stahlwilson1995}.\footnote{\citet{crawfordcostagomesiriberri2013} survey the approach.} Call the level-0 strategy profile from which the iteration starts the \emph{anchor}. The anchor and the distribution of levels are free parameters, and these have no counterpart in cursed rationalizability.

Do level-$k$ and cursed rationalizability predict different behavior? On the one hand, level-$k$ can accommodate cursed play. Call an anchor \emph{coarse} if its play is the same for every type; coarse anchors include the uniform-play specification standard in the literature. A level-1 player facing a coarse anchor holds a cursed conjecture, and with two players, level-1 play against coarse anchors generates exactly the first round of the cursed hierarchy---including every cursed-rationalizable strategy. A population of level-1 players, with anchors free to vary across players and types, can therefore generate any cursed-rationalizable strategy profile: from this perspective, cursed rationalizability is a refinement of level-$k$, selecting among predictions that the level-$k$ model can already generate. The freedom to set the anchor is what does the work: in the example above, entry is level-1 play only against anchors that conjecture participation with probability at least $\tfrac23$, and the uniform anchor will not do.

On the other hand, whatever the anchor, level-$k$ play survives $k$ rounds of the classical deletion: it lies in the $k$th set of the interim correlated rationalizability hierarchy. So whenever some cursed-rationalizable action is not interim correlated rationalizable, cursed play is distinct from level-$k$ play at sufficient depth. The contrast reflects where each theory puts the bias: In level-$k$, the bias is in the initial condition, and each step is computed by classical best response. By contrast, in cursed rationalizability, the initial condition is classical, and each step is computed by cursed response. In the example above, level~2 best-responds with correct inference to player 1's dominant strategy and stays out, from every anchor. Entry survives level-$k$ reasoning only at depth one; it survives cursed reasoning at every depth.

Disentangling cursed response from the equilibrium assumption sheds new light on laboratory data. It turns out that some experiments do not rely on the equilibrium assumption for identification. For example, in the seminal takeover game, buyers make take-it-or-leave-it offers to a seller who knows the value of the company \citep{samuelsonbazerman1985,holtsherman1994}. The company is worth $v$ to the seller and $1.5v$ to the buyer, with $v$ uniformly distributed on an interval $[X,X+R]$. \citet{holtsherman1994} found that, by varying the parameters $X$ and $R$, one can induce buyers to make offers that are too high or too low compared to the Bayesian Nash equilibrium prediction.\footnote{\citet{holtsherman1994} have a computer with a known strategy play the role of the seller. But we conjecture that their results are robust to replacing the computer with a human subject. Cursed rationalizability provides an argument for why they would be entitled to do so.}  In this setting, interim correlated rationalizability does not rule out any strategy profile. If the seller always rejects, then any offer is a best response for the buyer. And if the buyer's offer equals the seller's value, then any acceptance threshold is a best response for the seller. Thus, we apply a cautious refinement, deleting in the first round any action that is not a best response to a full-support conjecture \citep{dekelfudenberg1990}. We apply the parallel refinement to cursed rationalizability. With this refinement, the headline results of \citet{holtsherman1994} are inconsistent with interim correlated rationalizability and directionally consistent with cursed rationalizability. We also apply cursed rationalizability to second-price common-value auctions \citep{averykagel1997} and to speculative betting games \citep{sonsinoerevgilat2002,sovik2009}, with similar results.

\subsection{Related literature.} In addition to cursed equilibrium, there are other solution concepts in which players respond to coarse or misspecified beliefs, including analogy-based expectation equilibrium \citep{jehiel2005, jehielkoessler2008}, behavioral equilibrium \citep{esponda2008}, and Berk--Nash equilibrium \citep{espondapouzo2016}. Two recent papers extend cursedness to dynamic games: sequential cursed equilibrium \citep{cohenli2026} and cursed sequential equilibrium \citep{fonglinpalfrey2025}. Each of these is an equilibrium theory: conjectures are correct, up to the theory's coarsening or misspecification. This paper separates the response rule from the equilibrium hypothesis. In contemporaneous work, \citet{eystergagnonbartschrabin2026} propose a definition of cursed rationalizability that is specific to bilateral trade. Their definition is distinct from the one studied here, and Appendix~\ref{app:egr} records a distinguishing example. For generic trading games, the cautious refinement of cursed rationalizability, which we define in Section~\ref{sec:experiments}, is equivalent to their definition, as we show in Appendix~\ref{app:egr}.

A second literature models initial play in games. Level-$k$ reasoning \citep{stahlwilson1994, stahlwilson1995, nagel1995} and cognitive hierarchy models \citep{camererhochong2004} start from a na\"{i}ve anchor and iterate best responses a bounded number of steps; \citet{crawfordcostagomesiriberri2013} survey the theory and evidence. \citet{crawfordiriberri2007} use level-$k$ models to explain the winner's curse in auctions. Section~\ref{sec:levelk} compares the two approaches formally.

Third, an empirical literature documents cursedness in the laboratory. The winner's curse appears in common-value auctions \citep{kagellevin1986}, in the takeover game \citep{samuelsonbazerman1985, ballbazermancarroll1991, holtsherman1994}, and in speculative betting \citep{sonsinoerevgilat2002, sovik2009}. Section~\ref{sec:experiments} revisits three of these designs through the lens of cursed rationalizability. \citet{kagel1995} and \citet{kagellevin2016} survey the auction evidence. Recent experiments find that the winner's curse is substantially due to failures in contingent thinking \citep{charnesslevin2009, martinezmarquinaniederlevespa2019, nagel2025}. 

\section{Definitions}

\label{sec:concept}

We work with finite Bayesian games in the formulation of \citet{dekelfudenbergmorris2007}: payoffs depend on actions and on a payoff-relevant state, and types carry information about the state.

A \textbf{finite Bayesian game} is a tuple $G=(A_{1},\dots,A_{N};\,\Theta;\,T_{1},\dots,T_{N};\,p;\,u_{1},\dots,u_{N})$: players $i\in\{1,\dots,N\}$; finite action sets $A_{i}$, with $A:=\prod_{i}A_{i}$ and $A_{-i}:=\prod_{j\neq i}A_{j}$; a finite set $\Theta$ of \textbf{payoff-relevant states};\footnote{The state cannot in general be replaced by its conditional expectation given the type profile: conjectures (Definition~\ref{def:icr}) may correlate opponents' actions with $\theta$ beyond what opponents' types carry, and this correlation can enlarge the interim correlated rationalizable set \citep[Example~1]{dekelfudenbergmorris2007}. The fully cursed hierarchy is unaffected by the replacement: a cursed projection (defined below) pairs the state with opponents' actions independently, so payoffs enter its best responses only through their conditional expectations given the player's type. The ICR hierarchy and the partially cursed hierarchies of Section~\ref{sec:chifam} are affected.} finite type sets $T_{i}$, with $T:=T_{1}\times\cdots\times T_{N}$ and $T_{-i}:=\prod_{j\neq i}T_{j}$; a common prior\footnote{The definitions do not depend on the common prior assumption, and the general results about cursed rationalizability and $\chi$-cursed rationalizability in Sections~\ref{sec:concept}--\ref{sec:super} would hold if each type $t_{i}$ instead holds an arbitrary posterior on $\Theta\times T_{-i}$. We assume a common prior to make the comparison with cursed equilibrium straightforward. The particular conclusions of Examples \ref{ex:disjoint} and \ref{ex:pennies}, including Proposition \ref{prop:pennies}, rely on the posteriors and information structures specified in those examples, and need not hold under arbitrary type-specific posteriors. The same is true of the predictions for the experimental designs in Section \ref{sec:experiments}.} $p\in\Delta(\Theta\times T)$ with $p(t_{i})>0$ for every $i$ and $t_{i}\in T_{i}$; and payoffs $u_{i}:A\times\Theta\to\mathbb{R}$. Observe that types carry information, not payoffs. This is without loss of generality. If $u_i$ takes the type profile $t$ as an argument, then we can enrich the state space to $\Theta':=\Theta\times T$, concentrate the prior on the event that the state's type coordinates equal the realized types, and modify $u_i$ so that it depends on only the enriched state $\theta'$ and not on the type profile $t$. In particular, a \textbf{private-value game}---payoffs that depend only on $(a,t_{i})$---is the case $\Theta=T$ with $u_{i}$ reading the $i$-th coordinate; we write such payoffs directly as $u_{i}(a;t_{i})$.

A \textbf{conjecture} for $t_{i}$ is a distribution $\mu\in\Delta(\Theta\times T_{-i}\times A_{-i})$ with $\marg_{\Theta\times T_{-i}}\mu=p(\cdot,\cdot\mid t_{i})$: a joint belief about the state, opponents' types, and opponents' actions that agrees with type $t_i$'s posterior over $\Theta \times T_{-i}$. For a conjecture $\mu$ for $t_{i}$, its \textbf{cursed projection} is 
\[
\mu^{\mathrm{c}}:=p(\cdot,\cdot\mid t_{i})\otimes\marg_{A_{-i}}\mu,
\]
the product of the player's posterior over $\Theta \times T_{-i}$ and the conjecture's distribution over opponents' actions: the projection keeps both marginals of $\mu$ and severs the joint. A strategy for player $j$ is denoted $\sigma_{j}:T_{j}\to\Delta(A_{j})$. Given a strategy profile $\sigma=(\sigma_{j})_{j \in \{1,\ldots,N\}}$, the \textbf{true conjecture} induced by $p$ and $\sigma_{-i}$ for $t_{i}$ draws opponents' types from the posterior and their actions from their strategies: $\mu(\theta,t_{-i},a_{-i}):=p(\theta,t_{-i}\mid t_{i})\prod_{j\neq i}\sigma_{j}(a_{j}\mid t_{j})$.

Action $a_{i}$ is a \textbf{best response} to a conjecture $\mu$ for $t_{i}$ if $a_{i}\in\arg\max_{a_{i}'\in A_{i}}\mathbb{E}_{\mu}\big[u_{i}\big((a_{i}',a_{-i}),\theta\big)\big]$. Similarly, $a_{i}$ is a \textbf{cursed response} to $\mu$ if it is a best response to the cursed projection $\mu^c$, that is $a_{i}\in\arg\max_{a_{i}'\in A_{i}}\mathbb{E}_{\mu^{\mathrm{c}}}\big[u_{i}\big((a_{i}',a_{-i}),\theta\big)\big]$.

\begin{definition}[Fully cursed equilibrium]\label{def:fce} A strategy profile $\sigma^{*}=(\sigma_{i}^{*})_{i}$ is a \textbf{fully cursed equilibrium (FCE)} if for every $i$ and $t_{i}$, every $a_{i}\in\supp\sigma_{i}^{*}(\cdot\mid t_{i})$ is a cursed response to the true conjecture $\mu_{t_{i}}^{*}$ induced by $p$ and $\sigma_{-i}^{*}$. \end{definition}

This agrees with \citeauthor{eysterrabin2005}'s formulation: the true conjecture's action marginal is the average of opponents' equilibrium play given $t_{i}$, and the projection evaluates each action against that average, treating it as independent of the state.

For our purposes, the classical benchmark is interim correlated rationalizability (ICR) \citep{dekelfudenbergmorris2007}:

\begin{definition}[Interim correlated rationalizability]\label{def:icr} For a finite Bayesian game, set $R^{t_{i}}(0):=A_{i}$ and, for $k\geq1$: $a_{i}\in R^{t_{i}}(k)$ iff $a_{i}\in R^{t_{i}}(k-1)$ and there is a distribution $\mu\in\Delta(\Theta\times T_{-i}\times A_{-i})$ such that 
\begin{enumerate}
\item[(1)] $\marg_{\Theta\times T_{-i}}\mu=p(\cdot,\cdot\mid t_{i})$; 
\item[(2)] $\mu(\theta,t_{-i},a_{-i})>0$ implies $a_{j}\in R^{t_{j}}(k-1)$ for every $j\neq i$; 
\item[(3)] $a_{i}$ is a best response to $\mu$. 
\end{enumerate}
The \textbf{interim correlated rationalizable set} of type $t_{i}$ is the limit $R^{t_{i}}(\infty):=\bigcap_{k\geq0}R^{t_{i}}(k)$. \end{definition}

Clause (1) says $\mu$ is a conjecture for $t_{i}$; clause (2) licenses the conjecture---each opponent type is paired only with actions that survive for that type.

Cursed rationalizability replaces best response in clause (3) with cursed response.

\begin{definition}[Cursed rationalizability]\label{def:typewise} Set $C^{t_{i}}(0):=A_{i}$ and, for $k\geq1$: $a_{i}\in C^{t_{i}}(k)$ iff $a_{i}\in C^{t_{i}}(k-1)$ and there is a distribution $\mu\in\Delta(\Theta\times T_{-i}\times A_{-i})$ such that 
\begin{enumerate}
\item[(1)] $\marg_{\Theta\times T_{-i}}\mu=p(\cdot,\cdot\mid t_{i})$; 
\item[(2)] $\mu(\theta,t_{-i},a_{-i})>0$ implies $a_{j}\in C^{t_{j}}(k-1)$ for every $j\neq i$; 
\item[(3$'$)] $a_{i}$ is a cursed response to $\mu$. 
\end{enumerate}
The \textbf{cursed-rationalizable set} of type $t_{i}$ is the limit $C^{t_{i}}(\infty):=\bigcap_{k\geq0}C^{t_{i}}(k)$. \end{definition}

Definitions~\ref{def:icr} and~\ref{def:typewise} differ only in the third clause. Throughout, $C$ and $R$ denote the type-indexed families $(C^{t_{i}})$ and $(R^{t_{i}})$, with stage arguments as needed. 

One interpretation of cursed equilibrium is that players literally hold incorrect beliefs about their opponents. Specifically, ``each player incorrectly believes that with positive probability each profile of types of the other players plays the same mixed action profile that corresponds to their average distribution of actions, rather than their true, type-specific action profile" \citep[p.~1624]{eysterrabin2005}. A recent experiment by \citet{nagel2025} finds evidence against the literal-beliefs interpretation of cursed equilibrium, by having subjects bid against computer players that mimic the subject's own past behavior.

Definition \ref{def:typewise} departs from the literal-beliefs interpretation. It treats cursedness as a pure inference error: the player takes account of each opponent type's incentives in forecasting play, as in clause (2), but neglects the correlation between the opponent's action and the state in computing her own best response ($3'$). For example, in a common-value auction, a cursed player understands that when her own signal is high, her opponents' signals are probably also high. She also understands that high opponent signals lead to high opponent bids. She just forgets that those bids are relevant to her own evaluation of the object.

\begin{lemma}[Basic properties]\label{obs:basic} Each $C^{t_{i}}(k)$ is nonempty and decreasing in $k$, and the recursion stabilizes after finitely many stages. \end{lemma}
\begin{proof}
Monotonicity and stabilization are by construction and finiteness. Nonemptiness: given nonempty stage-$(k-1)$ sets, select $x_{j}^{t_{j}}\in C^{t_{j}}(k-1)$ for every $(j,t_{j})$; the conjecture $\mu(\theta,t_{-i},a_{-i}):=p(\theta,t_{-i}\mid t_{i})\,\mathbf{1}\{a_{j}=x_{j}^{t_{j}}\ \forall j\neq i\}$ satisfies clauses (1) and (2) at every stage $m\leq k$, since survivor sets are nested, so any cursed response $b$ to $\mu$ satisfies $b\in C^{t_{i}}(m)$ for every $m\leq k$, by induction on $m$. 
\end{proof}

We return to the example from the introduction that shows that $C$ and $R$ can be disjoint, with notation in hand.

\begin{example}[The entry game]
\label{ex:disjoint} Two states, good and bad, $\Theta=\{g,b\}$, with $p(g)=\tfrac{3}{4}$; player 1 observes the state ($T_{1}=\{g,b\}$, her type equal to the state) and chooses whether to participate, $a_{1}\in\{0,1\}$, with payoff $a_{1}$ in state $g$ and $-a_{1}$ in state $b$, so type $g$ participates and type $b$ does not, strictly; after one round these are the types' surviving actions under both hierarchies. Player 2, with a single type, chooses whether to enter, $a_{2}\in\{0,1\}$: staying out ($a_{2}=0$) pays $0$, and entry ($a_{2}=1$) pays $1$ if player 1 participates, regardless of the state, and $-1$ (state $g$) or $-5$ (state $b$) if she does not. At stage two, licensing forces player $2$'s conjecture to the true type-contingent play. Under clause (3), the conjecture pairs $g$ with $1$ and $b$ with $0$, entry is worth $\tfrac{3}{4}(1)+\tfrac{1}{4}(-5)=-\tfrac{1}{2}<0$, so $R^{t_{2}}(2) =\{0\} = R^{t_{2}}(\infty)$. The classical player stays out. By contrast, under clause (3$'$), only the conjecture's action marginal matters. Averaging over her posterior on the state, entry is worth $-2$ against $a_{1}=0$ and $1$ against $a_{1}=1$, so a conjecture with action marginal $\nu(a_{1}=1)=q$ values entry at $-2+3q$: entry is a cursed response iff $q\geq\tfrac{2}{3}$. The true play aggregates to $q=\tfrac{3}{4}$ and entry is uniquely optimal, so $C^{t_{2}}(2)=\{1\} =C^{t_{2}}(\infty)$ and the cursed player enters. 
\end{example}

Cursedness has an equivalent description as classical reasoning about a different game, defined as follows. For $t_{i}\in T_{i}$, the \textbf{average payoff} is 
\[
\bar{u}_{i}(a;t_{i}):=\sum_{\theta}p(\theta\mid t_{i})\,u_{i}(a,\theta),\qquad a\in A.
\]

\begin{definition}[The virtual game]\label{def:avggame} For a finite Bayesian game $G$, the \textbf{virtual game} $\overline{G}$ is the Bayesian game with the same players, actions, types, and prior, but with private-value payoffs $\bar{u}_{i}(a;t_{i})$. \end{definition}

In the virtual game, player $i$'s payoffs are pinned down by her own type and the action profile. Thus, she has no need to account for the information link between opponent actions $a_{-i}$ and the state $\theta$ when deciding what to do.

\begin{proposition}[\citealp{eysterrabin2005}]\label{obs:fcebne} $\sigma^{*}$ is a fully cursed equilibrium of $G$ if and only if it is a Bayesian Nash equilibrium of $\overline{G}$. \end{proposition}

We now state the same equivalence, but for cursed rationalizability.

\begin{proposition}[The dictionary]\label{prop:avgdict} For every finite Bayesian game $G$ and every $i$, $t_{i}$, $k$: the cursed hierarchy of $G$ is the ICR hierarchy of the virtual game, $C^{t_{i}}(k)\,[G]=R^{t_{i}}(k)\,[\overline{G}]$. \end{proposition}
This is a special case of Corollary~\ref{prop:chidict}, which follows shortly.

Propositions \ref{obs:fcebne} and \ref{prop:avgdict} show that a cursed player is a {classical} reasoner about the wrong game---the private-value game in which the state has been integrated out of payoffs,  so that each player can evaluate outcomes without conditioning on opponent behavior.

From this perspective, we can interpret cursed response as a misspecified mental model\footnote{\citet{spiegler2016} proposed the seminal theory of misspecified causal models. See \citet{spiegler2020} for a review.}: A cursed player understands that opponent types $t_{-i}$ are linked to their actions $a_{-i}$; but when evaluating her own payoffs, she omits the link between $t_{-i}$ and $\theta$. The player mistakenly treats an interdependent-values game as a private-values game, applying heuristics that would be entirely rational under private values. For example, in an auction, she considers how her own signal affects her value for the object, and how her opponent's signal affects his value, but treats these as separate problems.

\subsection{Partial cursedness}

\label{sec:chifam}

The definitions and results generalize naturally to partial cursedness. For parameter $\chi\in[0,1]$, say that action $a_{i}$ is a \textbf{$\chi$-cursed response} to a conjecture $\mu$ for $t_{i}$ if 
\[a_{i}\in\arg\max_{a_{i}'\in A_{i}}\mathbb{E}_{(1-\chi)\mu+\chi\mu^{\mathrm{c}}}\big[u_{i}\big((a_{i}',a_{-i}),\theta\big)\big].\]
Observe that a $0$-cursed response is a best response, and a $1$-cursed response is a cursed response.

A \textbf{$\chi$-cursed equilibrium} \citep{eysterrabin2005} is a profile $\sigma^{*}$ such that each $a_{i}\in\supp\sigma_{i}^{*}(\cdot\mid t_{i})$ is a $\chi$-cursed response to the true conjecture $\mu_{t_{i}}^{*}$ induced by $p$ and $\sigma_{-i}^{*}$.

\begin{definition}[$\chi$-cursed rationalizability]\label{def:chitw} For $\chi\in[0,1]$, set $C_{\chi}^{t_{i}}(0):=A_{i}$ and, for $k\geq1$: $a_{i}\in C_{\chi}^{t_{i}}(k)$ iff $a_{i}\in C_{\chi}^{t_{i}}(k-1)$ and there is a distribution $\mu\in\Delta(\Theta\times T_{-i}\times A_{-i})$ such that 
\begin{enumerate}
\item[(1)] $\marg_{\Theta\times T_{-i}}\mu=p(\cdot,\cdot\mid t_{i})$; 
\item[(2)] $\mu(\theta,t_{-i},a_{-i})>0$ implies $a_{j}\in C_{\chi}^{t_{j}}(k-1)$ for every $j\neq i$; 
\item[(3$_{\chi}$)] $a_{i}$ is a $\chi$-cursed response to $\mu$. 
\end{enumerate}
The \textbf{$\chi$-cursed-rationalizable set} of type $t_{i}$ is the limit $C_{\chi}^{t_{i}}(\infty):=\bigcap_{k\geq0}C_{\chi}^{t_{i}}(k)$. \end{definition}

The conjecture $(1-\chi)\mu+\chi\mu^{\mathrm{c}}$ interpolates between the conjectured joint over $(\theta,t_{-i},a_{-i})$ and the product of its two marginals---the state--type posterior and the action distribution. Setting $\chi=0$ yields interim correlated rationalizability and $\chi = 1$ yields cursed rationalizability. Observe that every $\chi$-cursed equilibrium has support in the $\chi$-cursed-rationalizable set.

The proof of Lemma~\ref{obs:basic} extends \textit{mutatis mutandis} to arbitrary $C_{\chi}$.

\citet{eysterrabin2005} define a $\chi$-virtual game as follows: For $\chi\in[0,1]$, let $G_{\chi}$ denote the Bayesian game with the same players, actions, types, and prior as $G$, and payoffs 
\[
u_{i}^{\chi}(a,\theta;t_{i})\;:=\;(1-\chi)\,u_{i}(a,\theta)\;+\;\chi\,\bar{u}_{i}(a;t_{i}).
\]

\begin{proposition}[Response equivalence]\label{prop:chiresp} For every $\chi\in[0,1]$, every type $t_{i}$, every conjecture $\mu$ for $t_{i}$, and every action $a_{i}$: $a_{i}$ is a $\chi$-cursed response to $\mu$ in $G$ if and only if $a_{i}$ is a best response to $\mu$ in $G_{\chi}$. \end{proposition}
\begin{proof}
Let us define the \textbf{cursed evaluation} of a distribution $\nu\in\Delta(A_{-i})$ as the expected utility under the \emph{product} of the state posterior and the action distribution,
\begin{equation}\label{eq:cursedeval}
V_{i}(a_{i};\nu,t_{i})\;:=\;\mathbb{E}_{p(\cdot\mid t_{i})\otimes\nu}\big[u_{i}\big((a_{i},a_{-i}),\theta\big)\big]\;=\;\sum_{a_{-i}}\nu(a_{-i})\,\bar{u}_{i}\big((a_{i},a_{-i});t_{i}\big).
\end{equation}
Since $\mu^{\mathrm{c}}=p(\cdot,\cdot\mid t_{i})\otimes\marg_{A_{-i}}\mu$, for any conjecture $\mu$ for $t_{i}$ and any $a_{i}'$, the cursed-response objective is $\mathbb{E}_{\mu^{\mathrm{c}}}\big[u_{i}\big((a_{i}',a_{-i}),\theta\big)\big]=V_{i}\big(a_{i}';\marg_{A_{-i}}\mu,t_{i}\big)=\mathbb{E}_{\mu}\big[\bar{u}_{i}\big((a_{i}',a_{-i});t_{i}\big)\big]$. Hence
\[
\mathbb{E}_{(1-\chi)\mu+\chi\mu^{\mathrm{c}}}\big[u_{i}\big]=\mathbb{E}_{\mu}\big[(1-\chi)u_{i}+\chi\bar{u}_{i}\big]=\mathbb{E}_{\mu}\big[u_{i}^{\chi}\big]:
\]
the $\chi$-belief evaluation of $\mu$ in $G$ is the classical evaluation of $\mu$ in $G_{\chi}$. The two maximization problems have the same feasible set $A_{i}$ and the same objective, so they have the same solutions. \end{proof}

\begin{corollary}[\citealp{eysterrabin2005}]\label{obs:chicebne} For $\chi\in[0,1]$, $\sigma^{*}$ is a $\chi$-cursed equilibrium of $G$ if and only if it is a Bayesian Nash equilibrium of $G_{\chi}$. \end{corollary}

\begin{corollary}[The $\chi$-dictionary]\label{prop:chidict} For every $\chi\in[0,1]$: $C_{\chi}(G)=R(G_{\chi})$, stage by stage. \end{corollary}

Partially cursed response seems strange, if we interpret an algorithm for solving the model as a literal mental process: Each player $i$ considers the joint distribution of $a_{-i}$ and $t_{-i}$, but then throws away some of that information by mixing the joint distribution with its marginals.

In light of Proposition \ref{prop:chiresp}, our preferred interpretation is that a partially cursed player leans too much on her private information when evaluating her own payoffs. At $\chi = 1$, she regards her type as a sufficient statistic. For $0 < \chi < 1$, the $\chi$-cursed player puts too much weight on her own type and too little on her opponents' types, which is captured by the $\chi$-virtual game $G_\chi$.

\subsection{Partial correlation}

\label{sec:pc}

We next consider a variation on the classical benchmark, that pertains especially to laboratory settings. A conjecture in Definition~\ref{def:icr} may correlate opponents' actions with the state directly, beyond what opponents' types carry. That is, the conjectured play at a given type profile may itself covary with $\theta$. This channel matters for behavior: in Example~2 of \citet{dekelfudenbergmorris2007}, speculative betting is permitted by ICR through conjectures that correlate the opponent's bet with the state in this way.

A natural refinement is to forbid such correlation. Call a conjecture $\mu$ \textbf{partially correlated} if $a_{-i}$ and $\theta$ are independent conditional on $t_{-i}$---equivalently, if $\mu(\theta,t_{-i},a_{-i})=p(\theta,t_{-i}\mid t_{i})\,\sigma_{-i}(a_{-i}\mid t_{-i})$ for some correlated conjectured strategy $\sigma_{-i}:T_{-i}\to\Delta(A_{-i})$. \textbf{Interim partially correlated rationalizability (IPCR)} is Definition~\ref{def:icr} but allowing only partially correlated conjectures; the concept was sketched by \citet{dekelfudenbergmorris2007} and formalized by \citet{tang2015}, with \citet{elypeski2006} studying the fully independent two-player case. When the state is a coin flip or a random number generator---as in a typical laboratory experiment---behavior cannot plausibly covary with the state except through information, so IPCR is arguably the right classical benchmark.

Cursed rationalizability is indifferent to the choice of benchmark: clause (3$'$) evaluates a conjecture only through its action marginal, and a conjecture and its partially correlated projection share one, so restricting Definition~\ref{def:typewise} to partially correlated conjectures changes nothing. Through the dictionary this is immediate---in the virtual game the state is gone from payoffs, so correlation with the state is payoff-irrelevant, and the ICR and IPCR hierarchies of $\overline{G}$ coincide.

The $\chi$-family, by contrast, is sensitive to the benchmark at $\chi<1$: Definition~\ref{def:chitw} with partially correlated conjectures defines \textbf{$\chi$-cursed partially-correlated rationalizability} ($\chi$-CPCR, written $C_{\chi}^{\mathrm{pc}}$), whose $\chi=0$ endpoint is the IPCR hierarchy rather than the ICR hierarchy, and whose $\chi=1$ endpoint is $C$, by the same projection argument. The dictionary respects the restriction: the evaluation identity in the proof of Proposition~\ref{prop:chiresp} holds conjecture by conjecture, and partial correlation is a property of the conjecture alone, so $C_{\chi}^{\mathrm{pc}}(G)$ is the IPCR hierarchy of $G_{\chi}$, stage by stage. Where the state is objectively random, this variation is the natural interpolation.

\subsection{Cursed rationalizability versus cursed equilibrium}\label{sec:vsce}

Is cursed rationalizability meaningfully different from cursed equilibrium? Example \ref{ex:disjoint} does not settle the question, since in that game cursed-rationalizability coincides with cursed equilibrium, and both are distinct from ICR. Similarly, in games of complete information, cursed rationalizability reduces to rationalizability and cursed equilibrium reduces to Nash equilibrium, but separation on such games does not shed light on the role of cursed response.

In this section, we show how cursed rationalizability permits behavior that is ruled out by cursed equilibrium and also ruled out by ICR. In the next example, for any $\chi \in [0,1]$, $\chi$-cursed equilibrium predicts that everyone takes the outside option. ICR makes the same prediction. Yet for every $\chi \geq \tfrac12$, the $\chi$-cursed rationalizable set retains every undominated action.

\begin{example}\label{ex:pennies}
There are two states, $\Theta = \{H, T\}$, equally likely. Player 1 (Row) observes the state ($T_1 = \{H, T\}$, the type reporting $\theta$). Player 2 (Column) has a single type. Both players choose from $\{h, t, o\}$. Each player calls a state or takes the safe action $o$ which pays $1$ for sure. Action $h$ calls state $H$, action $t$ state $T$. Payoffs are displayed with the usual convention:
\[
\begin{array}{c|c|c|c|}
\multicolumn{1}{c}{\theta = H} & \multicolumn{1}{c}{h} & \multicolumn{1}{c}{t} & \multicolumn{1}{c}{o} \\ \cline{2-4}
h & \pcell{4}{0} & \pcell{-4}{0} & \pcell{0}{1} \\ \cline{2-4}
t & \pcell{-8}{8} & \pcell{-8}{-8} & \pcell{-8}{1} \\ \cline{2-4}
o & \pcell{1}{0} & \pcell{1}{0} & \pcell{1}{1} \\ \cline{2-4}
\end{array}
\qquad\qquad
\begin{array}{c|c|c|c|}
\multicolumn{1}{c}{\theta = T} & \multicolumn{1}{c}{h} & \multicolumn{1}{c}{t} & \multicolumn{1}{c}{o} \\ \cline{2-4}
h & \pcell{-8}{-8} & \pcell{-8}{8} & \pcell{-8}{1} \\ \cline{2-4}
t & \pcell{-4}{0} & \pcell{4}{0} & \pcell{0}{1} \\ \cline{2-4}
o & \pcell{1}{0} & \pcell{1}{0} & \pcell{1}{1} \\ \cline{2-4}
\end{array}
\]
Observe that for Row, action $o$ strictly dominates $t$ for type $H$, and action $o$ strictly dominates $h$ for type $T$. So Row's decision reduces to calling `truthfully' or staying out. Calling truthfully pays $4$ if Column makes the same call, and $-4$ if Column makes the opposite call.

Column's calls pay zero whenever Row calls truthfully or stays out; against a false Row call, Column gets $-8$ for matching Row's call and $8$ for making the opposite call. Averaging Column's payoff over the states, action profile held fixed, gives him the following virtual payoffs:
\[
\begin{array}{c|ccc}
\bar{u}_2 & h & t & o \\ \hline
h & -4 & \phantom{-}4 & 1 \\
t & \phantom{-}4 & -4 & 1 \\
o & \phantom{-}0 & \phantom{-}0 & 1
\end{array}
\]
Observe that the virtual game for Column resembles matching pennies with an outside option: against $h$ he wants $t$, against $t$ he wants $h$, against $o$ he wants $o$. Row's incentives point the other way: calling truthfully is best when Column makes the same call. Four cursed responses close a cycle, drawn below: an arrow from $X$ to $Y$ means that $X$ is a cursed response to the conjecture that the opponent plays as in $Y$.
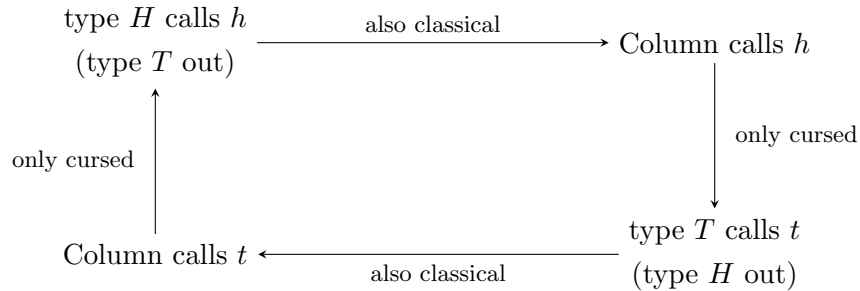
\begin{center}
\begin{tikzpicture}[>=stealth]
\node (Hh) at (0,0) [align=center] {type $H$ calls $h$\\ (type $T$ out)};
\node (Ch) at (7.4,0) {Column calls $h$};
\node (Ct) at (0,-2.8) {Column calls $t$};
\node (Tt) at (7.4,-2.8) [align=center] {type $T$ calls $t$\\ (type $H$ out)};
\draw[->] (Hh) -- node[above] {\footnotesize also classical} (Ch);
\draw[->] (Ch) -- node[right=4pt] {\footnotesize only cursed} (Tt);
\draw[->] (Tt) -- node[below] {\footnotesize also classical} (Ct);
\draw[->] (Ct) -- node[left=4pt] {\footnotesize only cursed} (Hh);
\end{tikzpicture}
\end{center}
The horizontal arrows are also classical best-responses; the vertical arrows are only cursed responses. Consider for example the vertical arrow on the right. If Column conjectures that type $T$ calls $t$ and type $H$ stays out, the action marginal $\tfrac12 t + \tfrac12 o$, paired with the uniform state posterior as if independent, values the call $h$ at $\tfrac12(4) + \tfrac12(0) = 2 > 1$. But its true value, correlation intact, is zero: the call $t$ comes only in state $T$, where Column's call $h$ wins nothing. In the depicted cycle, Row's two types hold conjectures that are not consistent with each other.
\end{example}

We state formally what each solution concept predicts in Example~\ref{ex:pennies}.

\begin{proposition}\label{prop:pennies}
In Example~\ref{ex:pennies}, list survivor families as (type $H$, type $T$, Column). For every $\chi \in [0,1]$:
\begin{enumerate}
\item[(1)] The unique $\chi$-cursed equilibrium is the all-$o$ profile.
\item[(2)] The hierarchy $C_\chi$ stabilizes after one round at $\big(\{h,o\}, \{t,o\}, \{h,t,o\}\big)$ if $\chi \geq \tfrac12$, and collapses, in three rounds, to $\big(\{o\}, \{o\}, \{o\}\big)$ if $\chi < \tfrac12$.
\item[(3)] Fully correlated and partially correlated conjectures license the same behavior, $C_{\chi} = C^{\mathrm{pc}}_{\chi}$ stage by stage. Moreover, ICR and IPCR both predict $\big(\{o\},\{o\},\{o\}\big)$.
\end{enumerate}
\end{proposition}

The proof of Proposition~\ref{prop:pennies} is in Appendix~\ref{app:proofs}.

To see why cursed equilibrium requires Column to stay out, observe that since calling falsely is strictly dominated, each Row type either stays out or calls truthfully. Suppose Row's $H$ type has a strictly higher probability of calling truthfully than the $T$ type. Then Column, who wishes to do the opposite of Row, must call $h$ with zero probability, which in turn means that the $H$ type strictly prefers to stay out, a contradiction. By symmetry, it also cannot be that Row's $T$ type has a strictly higher probability of calling truthfully than the $H$ type. It follows that both Row types call truthfully with equal probability, which implies that Column's expected virtual payoff from calling is $0$, so Column strictly prefers to stay out.

Equilibrium links all the players' conjectures through the same strategy profile. But in Example \ref{ex:pennies}, the conjectures that justify speculative betting cannot be simultaneously true. In this way, cursed rationalizability permits speculative bets that cursed equilibrium rules out, and that classical rationalizability rules out as well.

\section{Foundations}

\label{sec:found}

As one might expect, cursed rationalizability inherits some useful features of interim correlated rationalizability. In particular, $\chi$-cursed rationalizability is equivalent to the iterated deletion of interim strictly dominated strategies in the $\chi$-virtual game. Thus, $\chi$-cursed rationalizability is computationally tractable and the limit is invariant to deletion order.

We first state foundational results about $R$, and then extend these to $\chi$-cursed rationalizability.

For a family $X=(X_{j}^{t_{j}})_{j,t_{j}}$ of nonempty action sets and a type $t_{i}$, the \textbf{licensed-conjecture polytope} is 
\[
\mathcal{Q}^{t_{i}}(X):=\Big\{\mu\in\Delta(\Theta\times T_{-i}\times A_{-i})\;:\;\marg_{\Theta\times T_{-i}}\mu=p(\cdot,\cdot\mid t_{i}),\ \mu(\theta,t_{-i},a_{-i})>0\Rightarrow a_{j}\in X_{j}^{t_{j}}\ \forall j\neq i\Big\},
\]
the set of conjectures for $t_{i}$ that $X$ licenses. The set $\mathcal{Q}^{t_{i}}(X)$ is a nonempty compact convex polytope\footnote{The support restriction is a system of linear equations.}. Moreover, $\mathcal{Q}^{t_{i}}(X)$ is increasing in $X$ (coordinatewise), so it shrinks along the recursion. The definition of ICR (Definition~\ref{def:icr}) then reads: $a_{i}$ survives stage $k$ for $t_{i}$ iff $a_{i}$ is a best response to some $\mu\in\mathcal{Q}^{t_{i}}(R(k-1))$---call such a $\mu$ a \emph{stage-$k$ justifier}.\footnote{The definition's intersection clause, its requirement that $a_{i}\in R^{t_{i}}(k-1)$, is implied: the polytopes are nested along the recursion, so a stage-$k$ justifier witnesses justification at every stage $m\leq k$, and by induction the action already survives through stage $k-1$.}

Say that $a_{i}$ is \textbf{dominated for $t_{i}$ relative to $\mathcal{M}$}, for a set $\mathcal{M}$ of conjectures for $t_{i}$, if some $\sigma\in\Delta(A_{i})$ earns a strictly higher expected payoff than $a_{i}$ against every $\mu\in\mathcal{M}$. A family $X$ of nonempty action sets has the \textbf{best-response property} if every $a_{i}\in X_{i}^{t_{i}}$ is a best response to some $\mu\in\mathcal{Q}^{t_{i}}(X)$. The \textbf{$\chi$-cursed response property} is defined analogously.

\begin{proposition}[Foundations of ICR]\label{prop:icrfound} For every finite Bayesian game: 
\begin{enumerate}
\item[(i)] For every $k\geq1$: $a_{i}\in R^{t_{i}}(k)$ iff $a_{i}\in R^{t_{i}}(k-1)$ and $a_{i}$ is not dominated for $t_{i}$ relative to $\mathcal{Q}^{t_{i}}(R(k-1))$. 
\item[(ii)] $R(\infty)$ is computable by a number of linear programs polynomial in the size of the game. 
\item[(iii)] Call a \textbf{reduction} any decreasing sequence of families in which each step removes a nonempty set of triples $(i,t_{i},a_{i})$, each with $a_{i}$ dominated for $t_{i}$ relative to the current licensed-conjecture polytope, and call it maximal if it stops only when no such triple remains. Every maximal reduction terminates at $R(\infty)$ \citep[Theorem~1]{manili2024}. 
\item[(iv)] $R(\infty)$ has the best-response property, and it contains, coordinatewise, every family with the property \citep[Claim~3]{dekelfudenbergmorris2007}. In particular, every Bayesian Nash equilibrium $\sigma^{*}$ has $\supp\sigma_{i}^{*}(\cdot\mid t_{i})\subseteq R^{t_{i}}(\infty)$ for every $i,t_{i}$.
\end{enumerate}
\end{proposition}
\begin{proof}
(i) A best response to $\mu$ is beaten against $\mu$ by no mixture, so justifiability implies undominatedness. Conversely, consider the zero-sum game in which the maximizer chooses $\sigma\in\Delta(A_{i})$, the minimizer chooses $\mu\in\mathcal{Q}^{t_{i}}(R(k-1))$, and the payoff is the difference in expected payoffs between $\sigma$ and $a_{i}$. The payoff is bilinear and both domains are compact and convex, so the game has a value; if $a_{i}$ is undominated, the value is at most $0$, and a minimax-optimal $\mu^{*}$ holds every $a_{i}'\in A_{i}$ to at most $0$: $a_{i}$ is a best response to $\mu^{*}$, a stage-$k$ justifier. This is \citeauthor{pearce1984}'s lemma with the licensed-conjecture polytope in place of the full simplex. 

(ii) For fixed $(i,t_{i},a_{i},k)$, the existence of a stage-$k$ justifier is a feasibility program whose variables are the atoms of $\mu$ on the licensed support---at most $|\Theta|\cdot|T_{-i}|\cdot|A_{-i}|$ of them---and whose constraints are the marginal equalities, nonnegativity, and the $|A_{i}|-1$ optimality inequalities, each linear in $\mu$. Each stage solves at most $\sum_{i}|A_{i}|\,|T_{i}|$ such programs, and there are at most $\sum_{i}|A_{i}|\,|T_{i}|$ stages. 

(iii) and (iv) are from the cited results.
\end{proof}

By Corollary \ref{prop:chidict}, the hierarchy $C_\chi$ on game $G$ is equivalent to $R$ on the $\chi$-virtual game $G_\chi$. Moreover, the $\chi$-virtual game has the same size as $G$ and its payoffs can be computed from $G$ in polynomial time. Thus, Proposition \ref{prop:icrfound} yields the following corollary:

\begin{corollary}[Foundations of $\chi$-cursed rationalizability]\label{cor:chifound} For every finite Bayesian game and every $\chi\in[0,1]$: 
\begin{enumerate}
\item[(i)] For every $k\geq1$: $a_{i}\in C_{\chi}^{t_{i}}(k)$ iff $a_{i}\in C_{\chi}^{t_{i}}(k-1)$ and $a_{i}$ is not dominated for $t_{i}$ relative to $\mathcal{Q}^{t_{i}}(C_{\chi}(k-1))$, with payoffs evaluated in $G_{\chi}$. 
\item[(ii)] $C_{\chi}(\infty)$ is computable by a number of linear programs polynomial in the size of the game. 
\item[(iii)] Every maximal reduction---each step removing a nonempty set of triples $(i,t_{i},a_{i})$ with $a_{i}$ dominated for $t_{i}$ relative to the current licensed-conjecture polytope, with payoffs evaluated in $G_{\chi}$---terminates at $C_{\chi}(\infty)$. 
\item[(iv)] $C_{\chi}(\infty)$ has the $\chi$-cursed response property, and it contains, coordinatewise, every family with the property. In particular, every $\chi$-cursed equilibrium $\sigma^{*}$ has $\supp\sigma_{i}^{*}(\cdot\mid t_{i})\subseteq C_{\chi}^{t_{i}}(\infty)$ for every $i,t_{i}$.
\end{enumerate}
\end{corollary}

For fully cursed players ($\chi = 1$), the objective consults a conjecture only through its action marginal (the identity in the proof of Proposition~\ref{prop:chiresp}), so everything above can be run on the image polytope $\marg_{A_{-i}}\mathcal{Q}^{t_{i}}(X)\subseteq\Delta(A_{-i})$, the \emph{licensed aggregates}. We now state this formally.

\begin{lemma}[Strategy form]\label{lem:stratform} Cursed rationalizability depends only on the product form of conjectures: $a_{i}$ is a cursed response to a conjecture $\mu$ if and only if it is a cursed response to the strategy-form conjecture $\mu^{c}(\theta,t_{-i},a_{-i})=p(\theta,t_{-i}\mid t_{i})\,\sigma_{-i}(a_{-i}\mid t_{-i})$, where $\sigma_{-i}(\cdot\mid t_{-i}):=\marg_{A_{-i}}\mu(\cdot\mid t_{-i})$; and $\mu^{c}$ is licensed at every stage at which $\mu$ is. Equivalently, if and only if $a_{i}\in\arg\max_{a_{i}'\in A_{i}}V_{i}(a_{i}';\nu,t_{i})$ for the aggregate $\nu:=\sum_{t_{-i}}p(t_{-i}\mid t_{i})\,\sigma_{-i}(\cdot\mid t_{-i})$.
\end{lemma}
\begin{proof}
Given $\mu$ satisfying (1), (2), and (3$'$), and since $\sigma_{-i}(\cdot\mid t_{-i})=\marg_{A_{-i}}\mu(\cdot\mid t_{-i})$, licensing is preserved for each profile. The two conjectures share the action distribution $\marg_{A_{-i}}\mu$, which is all that clause (3$'$) consults, since $\mathbb{E}_{\mu^{\mathrm{c}}}[u_{i}]=V_{i}(\cdot\,;\marg_{A_{-i}}\mu,t_{i})$. The converse is immediate. 
\end{proof}

\section{Supermodular Bayesian games}

\label{sec:super}

Cursed rationalizability can be more permissive than cursed equilibrium. In the example of Section~\ref{sec:vsce}, the cursed rationalizable hierarchy retains every undominated action while the cursed equilibrium is unique. This section shows that strategic complements discipline that gap: the two concepts do not coincide, but they align at the extremes. At every degree $\chi$, the $\chi$-cursed rationalizable set for each type has a largest action and a smallest action, and those extremal profiles are $\chi$-cursed equilibria. Thus, equilibrium and common knowledge of $\chi$-cursed response yield the same bounds on behavior.

We state a definition of strategic complements for Bayesian games, requiring the \citet{milgromroberts1990} conditions to hold in every state.

\begin{definition}[Supermodular Bayesian game]\label{def:smod} $G$ is a \textbf{supermodular Bayesian game} if:
\begin{enumerate}
\item[(i)] each $A_{i}$ is a finite lattice, with $A_{-i}$ carrying the product order; and
\item[(ii)] each $u_{i}(\cdot,\theta)$ is \textbf{state-wise supermodular} in own action: $u_{i}((a_{i},a_{-i}),\theta)+u_{i}((b_{i},a_{-i}),\theta)\leq u_{i}((a_{i}\vee b_{i},a_{-i}),\theta)+u_{i}((a_{i}\wedge b_{i},a_{-i}),\theta)$ for all $a_{i},b_{i}$, and has \textbf{state-wise increasing differences}: for every $\theta$, $a_{i}\leq a_{i}'$, and $a_{-i}\leq a_{-i}'$, 
\[
u_{i}\big((a_{i}',a_{-i}),\theta\big)-u_{i}\big((a_{i},a_{-i}),\theta\big)\;\leq\;u_{i}\big((a_{i}',a_{-i}'),\theta\big)-u_{i}\big((a_{i},a_{-i}'),\theta\big).
\]
\end{enumerate}
\end{definition}

Next we show that supermodularity and increasing differences are well-behaved in supermodular Bayesian games.

\begin{lemma}[Inheritance]\label{lem:inherittw} In any supermodular Bayesian game $G$:
\begin{enumerate}
\item[(i)] every average payoff $\bar{u}_{i}(\cdot\,;t_{i})$ is supermodular in $a_{i}$ and has increasing differences in $(a_{i},a_{-i})$; hence every $V_{i}(\cdot\,;\nu,t_{i})$, defined in \eqref{eq:cursedeval}, is supermodular in $a_{i}$;
\item[(ii)] if $a_{i}'\geq a_{i}$ and $\nu'$ first-order stochastically dominates $\nu$ on $A_{-i}$ (that is, $\mathbb{E}_{\nu}f\leq\mathbb{E}_{\nu'}f$ for every weakly increasing $f:A_{-i}\to\mathbb{R}$, in the product order), then $V_{i}(a_{i}';\nu,t_{i})-V_{i}(a_{i};\nu,t_{i})\leq V_{i}(a_{i}';\nu',t_{i})-V_{i}(a_{i};\nu',t_{i})$;
\item[(iii)] the virtual game $\overline{G}$ is a supermodular Bayesian game; so is every convex combination of $G$ and $\overline{G}$, payoff by payoff.
\end{enumerate}
\end{lemma}
\begin{proof}
(i) and (iii): both properties are families of linear inequalities in payoffs, preserved by the convex combinations defining $\bar{u}_{i}$ and $V_{i}$. (ii): the map $a_{-i}\mapsto\bar{u}_{i}((a_{i}',a_{-i});t_{i})-\bar{u}_{i}((a_{i},a_{-i});t_{i})$ is increasing by (i), and first-order stochastic dominance raises the expectation of increasing functions. 
\end{proof}
We turn to characterizing the best responses to a specific opponent strategy. Consider a vector $x=(x_{j}^{t_{j}})_{j,t_{j}}$ of actions and a type $t_{i}$. Let $\mu_{x}$ correspond to the point conjecture of $x$ at type $t_{i}$: $\mu_{x}(\theta,t_{-i},a_{-i}):=p(\theta,t_{-i}\mid t_{i})\,\mathbf{1}\{a_{j}=x_{j}^{t_{j}}\ \forall j\neq i\}$, with aggregate $\bar{\nu}^{t_{i}}(x):=\marg_{A_{-i}}\mu_{x}=\sum_{t_{-i}}p(t_{-i}\mid t_{i})\,\delta_{(x_{j}^{t_{j}})_{j\neq i}}$, as in Lemma~\ref{lem:stratform}. With this notation, the $\chi$-expected utility of type $i$ against $x$ can be written as 
\begin{align*}
U_{i}^{\chi}(a_{i};x,t_{i})\;:=\;\mathbb{E}_{(1-\chi)\mu_{x}+\chi\mu_{x}^{\mathrm{c}}}\big[u_{i}\big((a_{i},a_{-i}),\theta\big)\big]\;
\\=\;(1-\chi)\sum_{(\theta,t_{-i})}p(\theta,t_{-i}\mid t_{i})\,u_{i}\big((a_{i},x(t_{-i})),\theta\big)\;+\;\chi\,V_{i}\big(a_{i};\bar{\nu}^{t_{i}}(x),t_{i}\big),
\end{align*}
where $x(t_{-i}):=(x_{j}^{t_{j}})_{j\neq i}$.

Each $U_{i}^{\chi}(\cdot\,;x,t_{i})$ is supermodular in $a_{i}$ as a convex combination of supermodular functions (state-wise supermodularity by Lemma \ref{lem:inherittw}(i)). Topkis's theorem implies that the set of maximizers $\arg\max_{a_{i}\in A_{i}}U_{i}^{\chi}(a_{i};x,t_{i})$ is a nonempty sublattice of $A_{i}$, with largest and smallest elements denoted by $\bar{b}_{\chi}^{t_{i}}(x)$ and $\underline{b}_{\chi}^{t_{i}}(x)$.

Let us define the sequences $\bar{x}_{\chi}^{t_{i},k},\underline{x}_{\chi}^{t_{i},k}$ as follows: The sequences begin at $\bar{x}_{\chi}^{t_{i},0}:=\max A_{i}$, $\underline{x}_{\chi}^{t_{i},0}:=\min A_{i}$, bounding the full range of actions, and proceed by $\bar{x}_{\chi}^{t_{i},k}:=\bar{b}_{\chi}^{t_{i}}(\bar{x}_{\chi}^{k-1})$, $\underline{x}_{\chi}^{t_{i},k}:=\underline{b}_{\chi}^{t_{i}}(\underline{x}_{\chi}^{k-1})$. The sequences are monotone; by Lemma ~\ref{lem:monotw}, $\bar{x}_{\chi}^{t_{i},k}$ is decreasing and $\underline{x}_{\chi}^{t_{i},k}$ increasing. Denote their limits by $\bar{x}_{\chi}^{t_{i}},\underline{x}_{\chi}^{t_{i}}$ (they stabilize by finiteness). For the endpoints we drop the subscript: let $\bar{x}^{t_{i},k}:=\bar{x}_{1}^{t_{i},k}$ (the cursed bounds) and $\bar{y}^{t_{i},k}:=\bar{x}_{0}^{t_{i},k}$ (the classical bounds). At $\chi=1$ the objective is $V_{i}(\cdot\,;\bar{\nu}^{t_{i}}(x),t_{i})$, and at $\chi=0$ it is the interim objective $U_{i}:=U_{i}^{0}$.

\begin{lemma}[Monotonicity]\label{lem:monotw} For every $\chi$, the differences of $U_{i}^{\chi}$ are increasing in $x$: if $a_{i}\leq a_{i}'$ and $x\leq x'$, then $U_{i}^{\chi}(a_{i}';x,t_{i})-U_{i}^{\chi}(a_{i};x,t_{i})\leq U_{i}^{\chi}(a_{i}';x',t_{i})-U_{i}^{\chi}(a_{i};x',t_{i})$. Thus, $\bar{b}_{\chi}^{t_{i}}$ and $\underline{b}_{\chi}^{t_{i}}$ are weakly increasing, the upper stage bounds decrease monotonically, and the lower stage bounds increase monotonically. \end{lemma}
\begin{proof}
Increasing differences in $(a_{i},x)$: $u_{i}\big((a_{i},x(t_{-i})),\theta\big)$ has increasing differences for any type profile ($x(t_{-i})\leq x'(t_{-i})$ coordinatewise), and the cursed term has them by Lemma~\ref{lem:inherittw}(ii), since $\bar{\nu}^{t_{i}}(x)\preceq\bar{\nu}^{t_{i}}(x')$ in first-order stochastic dominance; the convex combination inherits both. For the maps, let $x\leq x'$ and write $m:=\bar{b}_{\chi}^{t_{i}}(x)$, $m':=\bar{b}_{\chi}^{t_{i}}(x')$. Then 
\[
0\;\leq\;U_{i}^{\chi}(m;x)-U_{i}^{\chi}(m\wedge m';x)\;\leq\;U_{i}^{\chi}(m;x')-U_{i}^{\chi}(m\wedge m';x')\;\leq\;U_{i}^{\chi}(m\vee m';x')-U_{i}^{\chi}(m';x'),
\]
by optimality of $m$ at $x$, increasing differences (with $m\wedge m'\leq m$), and supermodularity. So $m\vee m'$ is optimal at $x'$, hence $m\vee m'\leq m'$, i.e., $m\leq m'$. Similarly, write $l:=\underline{b}_{\chi}^{t_{i}}(x)$, $l':=\underline{b}_{\chi}^{t_{i}}(x')$. Then 
\[
U_{i}^{\chi}(l;x)-U_{i}^{\chi}(l\wedge l';x)\;\leq\;U_{i}^{\chi}(l\vee l';x)-U_{i}^{\chi}(l';x)\;\leq\;U_{i}^{\chi}(l\vee l';x')-U_{i}^{\chi}(l';x')\;\leq\;0,
\]
by supermodularity, increasing differences (with $l'\leq l\vee l'$), and optimality of $l'$ at $x'$. So $l\wedge l'$ is optimal at $x$, hence $l\leq l\wedge l'$, i.e., $l\leq l'$. The bound sequences start at the extremes and the maps are monotone. 
\end{proof}
We next show that for all $k$, all the stage-$k$ $\chi$-cursed rationalizable actions lie in an interval whose endpoints are stage-$k$ $\chi$-cursed rationalizable actions. As $k$ increases, the endpoints of the intervals become closer. The interval stabilizes after finitely many steps.

\begin{theorem}[Interval theorem]\label{thm:typeinterval} In a finite supermodular Bayesian game, for every $\chi\in[0,1]$, $i$, $t_{i}$, $k$: $C_{\chi}^{t_{i}}(k)\subseteq\{a_{i}:\underline{x}_{\chi}^{t_{i},k}\leq a_{i}\leq\bar{x}_{\chi}^{t_{i},k}\}$, and $\underline{x}_{\chi}^{t_{i},k},\bar{x}_{\chi}^{t_{i},k}\in C_{\chi}^{t_{i}}(k)$. Hence $C_{\chi}^{t_{i}}(\infty)$ lies in the order interval $[\underline{x}_{\chi}^{t_{i}},\bar{x}_{\chi}^{t_{i}}]$; moreover, $\underline{x}_{\chi}^{t_{i}},\bar{x}_{\chi}^{t_{i}}\in C_{\chi}^{t_{i}}(\infty)$.
\end{theorem}
\begin{proof}
By induction on $k$. \emph{Attainment:} the point conjecture $\mu_{\bar{x}_{\chi}^{k-1}}$ satisfies clauses (1) and (2) of Definition~\ref{def:chitw} at every stage $m\leq k$ by the inductive hypothesis, and $\bar{x}_{\chi}^{t_{i},k}$ is a $\chi$-cursed response to it, since its $\chi$-belief evaluation is $U_{i}^{\chi}(\cdot\,;\bar{x}_{\chi}^{k-1},t_{i})$. Thus $\bar{x}_{\chi}^{t_{i},k}$ survives every stage $m\leq k$, as in Lemma~\ref{obs:basic}.

\emph{Deletion:} Let $a_{i}\in C_{\chi}^{t_{i}}(k)$ with a justifying conjecture $\mu$, and suppose $a_{i}\not\leq m:=\bar{x}_{\chi}^{t_{i},k}$, so the $\chi$-belief splits into its rational and cursed parts. For the rational, state-wise increasing differences atom by atom gives $\mathbb{E}_{\mu}[u_{i}(a_{i},\cdot)-u_{i}(a_{i}\wedge m,\cdot)]\leq U_{i}^{0}(a_{i};\bar{x}_{\chi}^{k-1},t_{i})-U_{i}^{0}(a_{i}\wedge m;\bar{x}_{\chi}^{k-1},t_{i})$; for the cursed part, $\marg_{A_{-i}}\mu\preceq\bar{\nu}^{t_{i}}(\bar{x}_{\chi}^{k-1})$ in first-order stochastic dominance, so Lemma~\ref{lem:inherittw}(ii) gives the same form for $V_{i}$. Weighting by $(1-\chi)$ and $\chi$, 
\begin{align*}
\mathbb{E}_{(1-\chi)\mu+\chi\mu^{\mathrm{c}}}\big[u_{i}(a_{i},\cdot)-u_{i}(a_{i}\wedge m,\cdot)\big] & \;\leq\;U_{i}^{\chi}\big(a_{i};\bar{x}_{\chi}^{k-1},t_{i}\big)-U_{i}^{\chi}\big(a_{i}\wedge m;\bar{x}_{\chi}^{k-1},t_{i}\big)\\
 & \;\leq\;U_{i}^{\chi}\big(a_{i}\vee m;\bar{x}_{\chi}^{k-1},t_{i}\big)-U_{i}^{\chi}\big(m;\bar{x}_{\chi}^{k-1},t_{i}\big)\\
 & \;<\;0,
\end{align*}
the second inequality by supermodularity and the third strictly because $a_{i}\vee m>m$ and $m$ is the largest maximizer of $U_{i}^{\chi}(\cdot\,;\bar{x}_{\chi}^{k-1},t_{i})$. So $a_{i}$ does not satisfy clause (3$_{\chi}$) of Definition~\ref{def:chitw} at $\mu$, a contradiction. The lower bound is symmetric.
\end{proof}
\begin{theorem}[Extremal profiles are $\chi$-cursed equilibria]\label{thm:typefce} For every $\chi\in[0,1]$, the pure strategy profiles $\bar{\sigma}_{\chi}:t_{i}\mapsto\bar{x}_{\chi}^{t_{i}}$ and $\underline{\sigma}_{\chi}:t_{i}\mapsto\underline{x}_{\chi}^{t_{i}}$ are $\chi$-cursed equilibria; hence, pure-strategy $\chi$-cursed equilibria exist in every finite supermodular Bayesian game. For every $\chi$-cursed equilibrium $\sigma$ and type $t_{i}$, $\supp\sigma_{i}(\cdot\mid t_{i})\subseteq[\underline{x}_{\chi}^{t_{i}},\bar{x}_{\chi}^{t_{i}}]$.
\end{theorem}

\begin{proof}
At the fixed point, $\bar{x}_{\chi}^{t_{i}}=\bar{b}_{\chi}^{t_{i}}(\bar{x}_{\chi})\in\arg\max_{a_{i}\in A_{i}}U_{i}^{\chi}(a_{i};\bar{x}_{\chi},t_{i})$. If every type $t_{j}$ of every opponent plays $\bar{x}_{\chi}^{t_{j}}$, the true conjecture induced by $p$ and the profile is exactly the point conjecture $\mu_{\bar{x}_{\chi}}$. The $\chi$-cursed equilibrium condition holds: a $\chi$-cursed response to $\mu_{\bar{x}_{\chi}}$ is precisely a maximizer of $U_{i}^{\chi}(\cdot\,;\bar{x}_{\chi},t_{i})$; symmetrically for $\underline{x}_{\chi}^{t_{i}}=\underline{b}_{\chi}^{t_{i}}(\underline{x}_{\chi})$. Containment is guaranteed by Corollary~\ref{cor:chifound}(iv) with Theorem~\ref{thm:typeinterval}, and $\bar{\sigma}_{\chi},\underline{\sigma}_{\chi}$ attain the bounds.
\end{proof}

Theorems \ref{thm:typeinterval} and \ref{thm:typefce} yield the corollary that the $\chi$-cursed-rationalizable set is a singleton exactly when $\chi$-cursed equilibrium is unique. 

\begin{corollary}[Dominance solvability at every degree]\label{cor:typeds} For every $\chi\in[0,1]$, the following are equivalent:
\begin{enumerate}
\item[(i)] $C_{\chi}^{t_{i}}(\infty)$ is a singleton for every $i,t_{i}$;
\item[(ii)] $\bar{x}_{\chi}=\underline{x}_{\chi}$;
\item[(iii)] the game has a unique $\chi$-cursed equilibrium.
\end{enumerate}
\end{corollary}
\begin{proof}
(i) $\Leftrightarrow$ (ii) by Theorem~\ref{thm:typeinterval}. (ii) $\Rightarrow$ (iii): every $\chi$-cursed equilibrium lies in the degenerate intervals, so it equals $\bar{\sigma}_{\chi}=\underline{\sigma}_{\chi}$, a $\chi$-cursed equilibrium by Theorem~\ref{thm:typefce}. (iii) $\Rightarrow$ (ii): $\bar{\sigma}_{\chi}$ and $\underline{\sigma}_{\chi}$ are both $\chi$-cursed equilibria. 
\end{proof}

\begin{corollary}[The classical mirror and the equilibrium gap]\label{thm:icrbne} At $\chi=0$, Theorems~\ref{thm:typeinterval} and~\ref{thm:typefce} read: the ICR hierarchy satisfies the interval theorem with the classical bounds $\bar{y}^{t_{i},k},\underline{y}^{t_{i},k}$, both attained; the profiles $t_{i}\mapsto\bar{y}^{t_{i}}$ and $t_{i}\mapsto\underline{y}^{t_{i}}$ are pure-strategy Bayesian Nash equilibria; and every BNE lies between them type by type. The gap between the cursed and classical boundary profiles is exactly the gap between extremal fully cursed and extremal Bayesian equilibria.
\end{corollary}

The $\chi=0$ result is an extension of Theorem 5 of \citet{milgromroberts1990} to ICR. By varying $\chi$, we find bounds for every degree of cursedness, attained by the theory's equilibria: fully cursed equilibria at $\chi=1$ and Bayesian Nash equilibria at $\chi=0$. The entry game of Example~\ref{ex:disjoint} is supermodular---entry's payoff is increasing in $a_{1}$ in each state, and staying out is constant---and player 2's endpoint intervals are disjoint: the cursed interval is $\{1\}$, the classical $\{0\}$.\footnote{The equilibrium half of the theorem holds more generally: \citet{vanzandt2010} proves existence of a greatest and a least interim Bayesian Nash equilibrium on arbitrary type spaces and compact metric lattice action sets, with no prior, by the same extremal iteration. What the theorem adds is the rationalizability half: containment in the interval at every degree, with bounds attained at every stage.}

\section{A comparison to level-\textit{k}}

\label{sec:levelk}

In this section, we compare the predictions of cursed rationalizability and level-$k$, which is a leading theory of initial play.

One obvious difference is that the predictions of level-$k$ depend on the anchor, and cursed rationalizability has no anchor. To make the parallel as close as possible, we study the level-$k$ predictions that can be generated by any anchor. Another obvious difference is that the level-$k$ prediction depends on the distribution of levels, and cursed rationalizability is defined by the limit of its operator (Definition~\ref{def:typewise}). To compare apples to apples, we compare $k$th-level play with the $k$th round of the cursed hierarchy.

An \textbf{anchor} is a profile $\sigma^{0}=(\sigma_{j}^{0})_{j}$ of mixed strategies $\sigma_{j}^{0}:T_{j}\to\Delta(A_{j})$. A \textbf{level-$k$ path} from $\sigma^{0}$ is any sequence $\Lambda_{1},\Lambda_{2},\dots$ of pure strategy profiles in which $\Lambda_{1}(t_{i})$ is a best response to the true conjecture induced by $p$ and $\sigma_{-i}^{0}$, and, for $k\geq2$, $\Lambda_{k}(t_{i})$ is a best response to the true conjecture induced by $p$ and $\Lambda_{k-1}$ \citep{crawfordiriberri2007}. In this dynamic the bias lives in the initial condition and the operator is classical: level $0$ is the anchor, and every step after it draws correct inferences. In cursed rationalizability the bias lives in the operator, and the initial condition is free.

From one perspective, cursed rationalizability refines level-$1$ play. Say that an anchor $\sigma^{0}$ is \textbf{coarse} if each $\sigma_{j}^{0}$ is type-independent, that is, $\sigma_{j}^{0}(t_j) = \sigma_{j}^{0}(t'_j)$ for all $t_j, t'_j$. Coarse anchors include the standard uniform-play specification.

\begin{proposition}\label{prop:coarseenv} In two-player games, the set of level-$1$ actions attainable across all coarse anchors is exactly $C^{t_{i}}(1)$; in general, it is contained in $C^{t_{i}}(1)$. \end{proposition}
\begin{proof}
Against a coarse anchor, the true conjecture is the product of the posterior and a type-independent action distribution, so it equals its cursed projection, and the classical and cursed objectives coincide. Every coarse anchor's aggregate, defined as in Lemma \ref{lem:stratform}, is a licensed stage-one conjecture, which gives the containment. Conversely, at stage one every conjectured strategy is licensed, so with two players the licensed aggregates are all of $\Delta(A_{j})$ (Lemma~\ref{lem:stratform}), each of them a coarse anchor. 
\end{proof}

\begin{corollary}\label{cor:refinelevel1}
    In two-player games, for any player $i$, type $t_i$, action $a_i$, and stage $k \geq 1$, if $a_i \in C^{t_i}(k)$ then there exists a coarse anchor $\sigma^0$ and a path from $\sigma^0$  such that $a_i$ is level-$1$ play by $t_i$.
\end{corollary}

From Corollary \ref{cor:refinelevel1}, we can interpret the cursed-rationalizable set as a refinement of level-$1$ play against coarse anchors. By Proposition \ref{prop:coarseenv}, it is a strict refinement whenever $C^{t_i}(1) \supsetneq C^{t_i}(\infty)$.

The results so far imply that, in two-player games, sufficiently flexible level-$k$ models can accommodate the cursed-rationalizable set. A population of level-1 players, with coarse anchors free to vary across players and types, can generate any strategy profile in $C(\infty)$.

Beyond level-$1$, the accommodation ends. Level-$k$ play is contained in the $k$th depth of the ICR hierarchy. Thus, any action that is cursed-rationalizable but not interim correlated rationalizable is, at sufficient depth, inconsistent with level-$k$ play regardless of the anchor. 

\begin{proposition}[Level-$k$ paths thread the classical hierarchy]\label{prop:levelkicr} For every anchor $\sigma^{0}$ and every level-$k$ path, we have $\Lambda_{k}(t_{i})\in R^{t_{i}}(k)$ for every $k$.
\end{proposition}
\begin{proof}
Set $\Lambda_{0}:=\sigma^{0}$ and induct on $k$: the conjecture pairing each $(\theta,t_{-i})$ with $\Lambda_{k-1}(t_{-i})$ has the posterior marginal on $\Theta\times T_{-i}$ and pairs each opponent type only with actions in $R^{t_{j}}(k-1)$ (at $k=1$ this is no constraint, since $R^{t_{j}}(0)=A_{j}$), so it is a conjecture licensed at stage $k$, and $\Lambda_{k}(t_{i})$ is a best response to it.
\end{proof}

The ICR hierarchy stabilizes after finitely many stages. Thus, Proposition \ref{prop:levelkicr} yields the following corollary.

\begin{corollary}
    For any anchor $\sigma^{0}$ and corresponding path $\Lambda_1, \Lambda_2, \dots$, and any type $t_i$, we have $\Lambda_k(t_i) \in R^{t_i}(\infty)$ for all sufficiently large $k$.
\end{corollary}

Every level-$k$ path, of every anchor, is ICR-consistent behavior at matching depth, and deep levels are classically rationalizable outright. The anchor in level-$k$ does not yield departures from $R$, but instead determines the path that level-$k$ traces through $R$.

Proposition \ref{prop:levelkicr} implies that if cursed rationalizability can be distinguished from the classical benchmark, then it can be distinguished at sufficient depth from level-$k$.

\begin{corollary}
    If $a_i \in C^{t_i}(\infty) \setminus R^{t_i}(\infty)$, then there exists finite $\underline{k}$ such that for any $k \geq \underline{k}$, any anchor $\sigma^{0}$ and corresponding path $\Lambda_1, \Lambda_2, \dots$, we have $a_i \neq \Lambda_{k}(t_{i})$.
\end{corollary}

For instance, in Example~\ref{ex:disjoint}, level-$k$ separates from the cursed hierarchy at depth $2$: we have $C^{t_{2}}(2)=\{1\}$ while $\Lambda_{2}(t_{2})=0$ for every anchor and every path.

The level-by-level comparison sheds light on how level-$k$ models contrast with cursed rationalizability. Once we fix an anchor, a level-$k$ model selects a path through the classical hierarchy $R$, and that path is unique except when actions are tied for best response. By contrast, cursed rationalizability defines a distinct hierarchy $C$ using cursed response instead of best response, and thus captures a different kind of non-equilibrium reasoning.

\section{Cursed rationalizability in experiments}

\label{sec:experiments}

A natural motivation for cursed rationalizability is cursed behavior in settings where experience is unavailable, making equilibrium play less plausible. Such behavior is observed in various laboratory games where cursedness appears on the initial rounds of play. This section studies the cursed hierarchy in three designs: the acquiring-a-company game \citep{samuelsonbazerman1985,holtsherman1994}, second-price common-value auctions \citep{averykagel1997}, and a zero-sum betting game \citep{sonsinoerevgilat2002,sovik2009}.

In all these designs, one player can veto meaningful strategic interaction; for instance, by rejecting all offers in the acquiring-a-company game, or by placing the maximum bid in a second-price auction. Thus, interim correlated rationalizability by itself has too little bite, because almost any action by the opponent can be rationalized when facing a veto. The same issue arises for cursed rationalizability.

We apply a cautious refinement of ICR and cursed rationalizability, requiring that conjectures must put positive probability on all opposing behavior. We apply caution only in the first round, following \citet{dekelfudenberg1990}, after which the ordinary recursions take over; call this the \textbf{cautious refinement}.\footnote{If we instead apply caution in every round, then the prediction can vary with deletion order \citep{marxswinkels1997}. Applying caution only in the first round prevents this issue. } Fix a family of nonempty type-indexed sets $X=(X_{j}^{t_{j}})_{j,t_{j}}$ and a type $t_{i}$.

\begin{definition}[Cautious justification]\label{def:cautious} Action $a_i$ is \textbf{cautiously cursed-justifiable} for $t_i$ given $X$ if $a_i$ is a cursed response to some conjecture $\mu$ licensed by $X$ whose conditional $\mu(\cdot \mid \theta, t_{-i})$ has full support on $\prod_{j \neq i} X_j^{t_j}$ for every $(\theta, t_{-i})$ with positive posterior. Action $a_{i}$ is \textbf{cautiously classically justifiable} for $t_{i}$ given $X$ if $a_{i}$ is a best response to some such conjecture. The refined recursions are denoted $\widetilde{C}$ (cursed) and $\widetilde{R}$ (classical). Caution is applied exactly once, at the first round, in the manner of \citet{dekelfudenberg1990}: $\widetilde{C}^{t_{i}}(1)$ contains the actions cautiously cursed-justifiable given the full family $(A_{j})_{j,t_{j}}$, and for $k\geq2$,
\[
\widetilde{C}^{t_{i}}(k):=\big\{a_{i}\in\widetilde{C}^{t_{i}}(k-1):a_{i}\text{ is a cursed response to some }\mu\in\mathcal{Q}^{t_{i}}(\widetilde{C}(k-1))\big\};
\]
likewise for $\widetilde{R}$, with cautious classical justifiability at the first round and best responses in place of cursed responses afterward. \end{definition}

The definition extends to partial cursedness: say that $a_{i}$ is cautiously $\chi$-justifiable for $t_{i}$ if $a_{i}$ is a $\chi$-cursed response to a conjecture with full-support conditionals, licensed as in Definition~\ref{def:cautious}, and run the $\chi$-recursion with the same design. This again requires only one cautious round, then Definition~\ref{def:chitw} and the same intersection convention; write $\widetilde{C}_{\chi}$ for it. Throughout, the classical benchmark is computed under the same refinement, so the comparison isolates the effect of cursedness. The one exception is the auctions of Section~\ref{sec:akexp}, where the refinement turns out to be insufficiently selective on the classical side and the benchmark is equilibrium play.

We find that in the experiment designs that follow, the inference bias typically associated with cursed behavior is predicted by cursed rationalizability with the cautious refinement. The further hypothesis of equilibrium coordination is not needed. Thus, the experiments are evidence in favor of the cursedness, even if one is skeptical that laboratory subjects are calibrated about their opponents' behavior.

These experiments do not distinguish cursed rationalizability from cursed equilibrium. In our view, there is no general answer about which theory is right. Whether players hold statistically correct beliefs about opponents' play likely depends on context and experience, both in the laboratory and in the real world.

The designs are computed with finite value grids and finite bid grids, as in the games the subjects actually faced. The paper's theorems are stated and proved for finite games, but nothing about the concepts is intrinsically finite, and the extension to games with interval type and action spaces is straightforward to state: conjectures become measurable joint distributions over states, opponents' types, and actions; the cursed projection again pairs the state with opponents' actions independently; and caution asks for conjectures with full support on surviving play. 

\subsection{The acquiring-a-company game}

\label{sec:hsexp}

In the acquiring-a-company game of \citet{samuelsonbazerman1985}, a buyer bids for a company whose value the seller knows and the buyer does not. The company is worth $v$ to the seller and $1.5v$ to the buyer. The buyer makes a single bid $b$, which is accepted if and only if $b\geq v$, in which case the buyer earns $1.5v-b$; otherwise both earn zero. \citet{holtsherman1994} run this experiment with $v$ uniform on $[X,X+R]$ and a simulated seller. We model the seller as a player whose strategy is a cutoff (accept iff $b\geq c$), with a cutoff that rejects every bid available, so that cutoff $c=v$ is weakly dominant. In the notation of Section~\ref{sec:concept}, the state is $v$: the seller observes it, the buyer does not. Three treatments (their Table~1) move the parameters: a \emph{winner's-curse} design ($X=1.50$, $R=4.50$), a \emph{loser's-curse} design ($X=0.50$, $R=0.50$), and a \emph{no-curse} design ($X=1.00$, $R=2.00$). Rational play earns the same expected amount in every treatment, and naïve play costs the same in the two cursed treatments, so the direction of the bias is the only thing that varies.

In this setting, cautious cursed rationalizability coincides with cursed equilibrium and with \citeauthor{holtsherman1994}'s \emph{naïve} benchmark, and cautious ICR coincides with their rational benchmark. The cautious round ensures that only the dominant cutoff survives for the seller, uniquely up to measure zero. Every licensed conjecture yields the acceptance probability $\tfrac{b-X}{R}$ on $[X,X+R]$. A cursed buyer therefore values bid $b$ at $\tfrac{b-X}{R}\,(1.5X+0.75R-b)$, replacing the conditional mean of $1.5v$ by its unconditional mean (Lemma~\ref{lem:stratform}). \citeauthor{holtsherman1994}'s naïve bid $1.25X+0.375R$ is the fully cursed equilibrium bid, and cautious cursed rationalizability predicts it uniquely. The buyer's true expected payoff is $\tfrac{b-X}{R}\cdot\tfrac{3X-b}{4}$, maximized at their \emph{rational} bid $2X$, which is what cautious ICR predicts. Design by design, the cursed and rational bids are: $3.56$ and $3.00$ in the winner's-curse design, so the cursed buyer overbids; $0.81$ and $1.00$ in the loser's-curse design, so she bids too little; and $2.00$ and $2.00$ in the no-curse design, equal by construction.

Holt and Sherman discretized this setting for the laboratory: their implemented design draws values in penny increments over the full range, $v\in\{X,\,X+0.01,\,\dots,\,X+R\}$, equally likely, and takes bids in penny increments.\footnote{Holt and Sherman's instructions describe the randomization twice: ``each number 1.00, 1.01, \dots, 3.00 will have an equal chance of being selected,'' and, in a question-and-answer summary, as a random number with uniform support on $\{0,.01,\ldots,.99\}$ times $R$, which would omit the top value and change the grid. We read the first description as operative.} This discretization creates penny-level ties in the equilibrium bids and widens the refined sets from points into short intervals: sellers are exactly indifferent between acceptance thresholds $v$ and $v+0.01$, so two thresholds can survive the cautious round, and each bid in the interval is a response to a conjecture resolving those ties in a different way. We have computed the predicted buyer bids of each solution concept for the discretized design, and state these in Figure~\ref{fig:hsexp}; the exact values are reported in Appendix~\ref{app:hsexp}.

\begin{figure}[h]
\begin{center}
\begin{tikzpicture}[x=3.8cm,y=1cm]
  \def\xmin{0.5}
  \node[anchor=east, font=\small] at (-0.02,1.75) {cursed};
  \node[anchor=east, font=\small] at (-0.02,1.05) {classical};

  \fill[cursedfill, rounded corners=1pt] ({0.76-\xmin},1.60) rectangle ({0.86-\xmin},1.90);
  \fill[cursedfill, rounded corners=1pt] ({1.90-\xmin},1.60) rectangle ({2.09-\xmin},1.90);
  \fill[cursedfill, rounded corners=1pt] ({3.42-\xmin},1.60) rectangle ({3.70-\xmin},1.90);
  \fill[cursedcol] ({0.81-\xmin},1.75) circle (2.4pt);
  \fill[cursedcol] ({1.995-\xmin},1.75) circle (2.4pt);
  \fill[cursedcol] ({3.56-\xmin},1.75) circle (2.4pt);

  \fill[classicalfill, rounded corners=1pt] ({0.86-\xmin},0.90) rectangle ({1.00-\xmin},1.20);
  \fill[classicalfill, rounded corners=1pt] ({1.80-\xmin},0.90) rectangle ({2.19-\xmin},1.20);
  \fill[classicalfill, rounded corners=1pt] ({2.76-\xmin},0.90) rectangle ({3.23-\xmin},1.20);
  \draw[classicalcol, thick, fill=white] ({0.995-\xmin},1.05) circle (2.4pt);
  \draw[classicalcol, thick, fill=white] ({1.995-\xmin},1.05) circle (2.4pt);
  \draw[classicalcol, thick, fill=white] ({2.995-\xmin},1.05) circle (2.4pt);

  \foreach \d in {0.74, 2.03, 3.78} {
    \draw[dashed, thick] ({\d-\xmin},0.55) -- ({\d-\xmin},2.20);
  }
  \node[font=\scriptsize\itshape, anchor=south] at ({0.74-\xmin},2.20) {mean bid $0.74$};
  \node[font=\scriptsize\itshape, anchor=south] at ({2.03-\xmin},2.20) {mean bid $2.03$};
  \node[font=\scriptsize\itshape, anchor=south] at ({3.78-\xmin},2.20) {mean bid $3.78$};

  \node[font=\small\bfseries, anchor=south] at ({0.88-\xmin},2.65) {Loser's curse};
  \node[font=\small\bfseries, anchor=south] at ({2.00-\xmin},2.65) {No curse};
  \node[font=\small\bfseries, anchor=south] at ({3.27-\xmin},2.65) {Winner's curse};

  \draw[->] (-0.05,0.30) -- ({4.05-\xmin},0.30) node[anchor=west, font=\scriptsize] {bid};
  \foreach \t in {0.5,1.0,1.5,2.0,2.5,3.0,3.5,4.0} {
    \draw ({\t-\xmin},0.25) -- ({\t-\xmin},0.35);
    \node[font=\scriptsize, anchor=north] at ({\t-\xmin},0.22) {\t0};
  }

  \begin{scope}[shift={(0,-0.9)}, x=1cm]
    \fill[cursedfill, rounded corners=1pt] (0.0,0) rectangle (0.55,0.22);
    \node[anchor=west, font=\scriptsize] at (0.65,0.11) {cursed rationalizability};
    \fill[classicalfill, rounded corners=1pt] (3.9,0) rectangle (4.45,0.22);
    \node[anchor=west, font=\scriptsize] at (4.55,0.11) {interim correlated rationalizability};
    \fill[cursedcol] (9.5,0.11) circle (2.1pt);
    \node[anchor=west, font=\scriptsize] at (9.65,0.11) {fully cursed eq.};
    \draw[classicalcol, thick, fill=white] (12.1,0.11) circle (2.1pt);
    \node[anchor=west, font=\scriptsize] at (12.25,0.11) {Bayesian Nash eq.};
  \end{scope}
\end{tikzpicture}
\end{center}
\caption{Predictions and data in the three treatments of \citet{holtsherman1994}. Equilibria are computed with the seller accepting at her threshold value. Rationalizability concepts are computed under the cautious refinement.}
\label{fig:hsexp}
\end{figure}
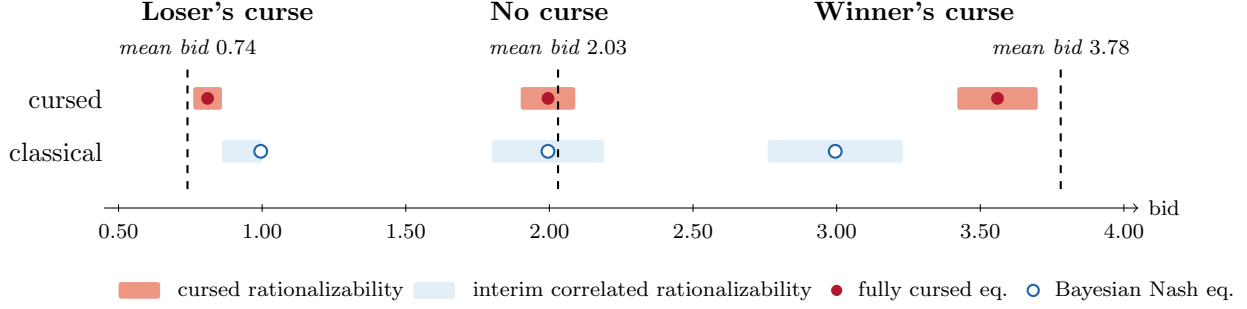

In the winner's curse and loser's curse treatments, the data align with the fully cursed predictions; those of cursed equilibrium and cautious cursed rationalizability. The observed bids are even higher than the cursed predictions in the winner's curse treatment, and lower than the cursed predictions in the loser's curse treatment. In the no-curse treatment, all four theories yield almost the same predictions, and are broadly consistent with the data.

A similar experiment was conducted by \citet{charnesslevin2009} as an individual-choice task. There, the value $v$ is uniform on $\{0,\dots,99\}$ and the seller is replaced by a card the bidder flips, with no other agent in any role. The overbidding persists there, despite the absence of strategic interaction. The overbidding is largely eliminated by a reframing of a two-value version as an explicit choice among lotteries, which removes the need for contingent reasoning. Part of the mechanism is thus a failure of contingent reasoning that is not specific to strategic interaction.

\subsection{Second-price common-value auctions}

\label{sec:akexp}

A classic setting for the winner's curse is second-price common value auctions; \citet{averykagel1997} tested a simple variant with only two bidders. Two signals $x,y$ are drawn independently and uniformly on $(1,4)$; the common value is $V=x+y$. In the \emph{private-value-advantage} treatment one bidder additionally receives a known $K=1$ upon winning. The state is the signal pair, each bidder observing her own coordinate.

The virtual game of {any} common-value second-price auction has private values $E[V\mid t_{i}]$, so bidding one's value is weakly dominant and every other bid is inadmissible. The cautious round alone solves the cursed problem and cautious cursed rationalizability predicts that each player bids their naïve expected value: $B_{\mathrm{EV}}(x)=x+2.5$ in the standard auction and, with the advantage, $x+3.5$ for the advantaged bidder and $y+2.5$ for the disadvantaged bidder. This is also the fully cursed equilibrium. Cautious ICR is far weaker: the cautious round deletes only the bids outside the range of possible values, leaving $[x+1,x+4]$ for signal $x$ (shifted up by $K$ for the advantaged bidder), and iteration then adds nothing. Because cautious ICR is insufficiently selective here, the classical benchmark we carry to the data is equilibrium play. In the symmetric treatment, the symmetric BNE bid is $B^{*}(x)=2x$.\footnote{Re-applying caution at every round selects $B^{*}$ in the standard treatment, but that procedure inherits the order-dependence of iterated admissibility \citep{marxswinkels1997}, and second-price common-value auctions in any case have families of asymmetric equilibria for which we make no general selection claim.} With the advantage, any well-behaved BNE leads to a win of the advantage bidder with probability one (\citet{averykagel1997} Theorem~2.2, after \citet{bikhchandani1988}). We now characterize cursed rationalizability and ICR for discretizations of the Avery--Kagel design.

\begin{proposition}\label{prop:akexp} Consider any discretization of the \citet{averykagel1997} design in which the signals are i.i.d.\ uniform on a finite set $S\subset[1,4]$ that contains $1$, $2.5$, and $4$ and is symmetric about $2.5$, and both bidders choose bids from a finite set $\mathcal{B}$ that contains every possible value $x+y+K_{i}$, where $K_{i}$ is bidder $i$'s private-value advantage ($K_{i}=1$ for the advantaged bidder in the private-value-advantage treatment and $K_{i}=0$ otherwise). For signal $x$, let $\underline{m}_{i}(x):=x+1+K_{i}$ and $\overline{m}_{i}(x):=x+4+K_{i}$ denote bidder $i$'s smallest and largest possible values. In both treatments: (i) without caution, both hierarchies retain every bid for every signal; (ii) the refined cursed hierarchy is dominance-solvable at exactly $B_{\mathrm{EV}}$, in a single round; (iii) the refined classical hierarchy retains exactly the bids in $[\underline{m}_{i}(x),\overline{m}_{i}(x)]$ for signal $x$ after the cautious round, and no further round deletes anything. \end{proposition}
The proof of Proposition~\ref{prop:akexp} is in Appendix~\ref{app:proofs}.

The refined cursed hierarchy and the BNE benchmark land exactly on the two bid functions \citet{averykagel1997} test, and the data favor the cursed one. In standard auctions their estimated bid function for inexperienced bidders is $B(x)=1.13x + 2.64$: they cannot reject the cursed bid function $B_{\mathrm{EV}}(x)=x+2.5$ ($p=0.21$) and they reject the symmetric BNE $B^*(x) = 2x$ ($p<0.01$). For experienced bidders the estimated bid function is $B(x)=1.34x + 1.99$. Both point predictions are rejected (at $p < 0.01$) but the estimates remain closer to $B_{\mathrm{EV}}$, with slow movement toward Nash. Cursed bidding is more prevalent for inexperienced bidders, which suggests that cursed rationalizability has a role to play in explaining the data.

In the advantage treatment, \citet{averykagel1997} find that the effect on bids is proportional rather than explosive: advantaged bidders win $62$--$71$ percent of auctions, not the $100$ percent predicted by BNE. By contrast, the refined cursed bids predict a win rate of about $78$ percent. These findings are broadly consistent with the refined cursed prediction of Proposition~\ref{prop:akexp}.

\subsection{Speculative betting}

\label{sec:sovikexp}

We study experimental data on a betting game with zero-sum bets from \citet{sovik2009}.\footnote{The betting game was introduced by \citet{sonsinoerevgilat2002} in an unpublished working paper.} There are four states, $A$, $B$, $C$, $D$, with equal priors $\tfrac{1}{4}$; player 1 observes the partition $\{A,B\},\{C,D\}$ and player 2 the partition $\{A\},\{B,C\},\{D\}$, so the pair of information sets uniquely determines the state. Players choose simultaneously whether to bet or take an outside option with payoff $1$. The bet is on if and only if both players bet. Payoffs from bets are displayed in Table~\ref{tab:sovik}. In the notation of Section~\ref{sec:concept}, the state is $\theta\in\{A,B,C,D\}$ and the types are the information sets.

\begin{table}[!htbp]
\caption{Payoffs and types for the betting game, weak-dominance design. Each state has prior probability $\tfrac{1}{4}$.}
\label{tab:sovik}
\begin{center}
\begin{tabular}{l|cccc}
 & $A$ & $B$ & $C$ & $D$ \\
\hline
bet & $32,\,-32$ & $-28,\,28$ & $20,\,-20$ & $-16,\,16$ \\
no bet & $1,\,1$ & $1,\,1$ & $1,\,1$ & $1,\,1$ \\
\hline
player 1 types & \multicolumn{2}{c|}{$\{A,B\}$} & \multicolumn{2}{c|}{$\{C,D\}$} \\
\cline{2-5}
player 2 types & \multicolumn{1}{c|}{$\{A\}$} & \multicolumn{2}{c|}{$\{B,C\}$} & \multicolumn{1}{c|}{$\{D\}$} \\
\cline{2-5}
\end{tabular}
\end{center}
\end{table}

Conditional on her own information, every type except $\{A\}$ expects to gain from the bet: the average betting payoffs are $+2$ for both of player 1's types, $+4$ for type $\{B,C\}$, and $+16$ for type $\{D\}$. However, iterated deletion of weakly dominated strategies implies no betting: player 2 declines at $\{A\}$; then a bet at $\{A,B\}$ is on only in state $B$ and player 1 declines; then 2 declines at $\{B,C\}$, and finally 1 declines at $\{C,D\}$. Following \citeauthor{sovik2009}, we call this the \emph{weak-dominance design}. We have computed each solution concept for this design, and state the results in Proposition~\ref{prop:sovikexp}.

\begin{proposition}\label{prop:sovikexp} 
In the weak-dominance design:
\begin{enumerate}
\item[(i)] Iterated deletion of weakly dominated strategies predicts no betting at every type except $\{D\}$, which bets.\footnote{Here we assume that all weakly dominated strategies are deleted at each round. For some orders of deletion, both actions survive for type $\{D\}$.}
\item[(ii)] The profile in which player 1 bets at both types and player 2 bets at $\{B,C\}$ and $\{D\}$ but not at $\{A\}$ is a fully cursed equilibrium; the bet is realized in states $B$, $C$, and $D$.
\item[(iii)] Cautious cursed rationalizability uniquely selects the profile in (ii).
\item[(iv)] Cautious classical rationalizability selects no betting at $\{A\}$, betting at $\{D\}$, and keeps both actions everywhere else.
\item[(v)] Without caution, neither hierarchy eliminates anything.
\item[(vi)] The cautious $\chi$-cursed hierarchy has three regimes: for $\chi<\tfrac{7}{8}$ it coincides with (iv); for $\tfrac{7}{8}\leq\chi<\tfrac{29}{30}$ it uniquely selects betting at $\{B,C\}$, $\{C,D\}$, and $\{D\}$ and no betting at $\{A\}$ or $\{A,B\}$; and for $\chi\geq\tfrac{29}{30}$ it uniquely selects the profile of (ii).
\end{enumerate}
\end{proposition}

Proposition \ref{prop:sovikexp} shows that equilibrium is not needed to deliver speculative bets in this setting. Cautious cursed rationalizability selects a fully cursed equilibrium with betting. By contrast, cautious classical rationalizability is more permissive, and is consistent with betting and also with no betting.

\begin{figure}[!htbp]
\begin{center}
\begin{tikzpicture}[x=2.2cm, y=0.038cm]
  \draw[->] (0.4,0) -- (0.4,108);
  \foreach \yy in {0,25,50,75,100} {
    \draw (0.35,\yy) -- (0.45,\yy);
    \node[font=\scriptsize, anchor=east] at (0.33,\yy) {\yy};
    \draw[gray!30, very thin] (0.45,\yy) -- (5.6,\yy);
  }
  \node[font=\scriptsize, rotate=90, anchor=south] at (-0.05,50) {betting rate (percent)};
  \draw (0.4,0) -- (5.6,0);

  \foreach \xx/\lab in {1/{$\{A\}$}, 2/{$\{A,B\}$}, 3/{$\{B,C\}$}, 4/{$\{C,D\}$}, 5/{$\{D\}$}} {
    \node[font=\small, anchor=north] at (\xx,-3) {\lab};
  }
  \node[font=\scriptsize\itshape, anchor=north] at (1,-14) {player 2};
  \node[font=\scriptsize\itshape, anchor=north] at (2,-14) {player 1};
  \node[font=\scriptsize\itshape, anchor=north] at (3,-14) {player 2};
  \node[font=\scriptsize\itshape, anchor=north] at (4,-14) {player 1};
  \node[font=\scriptsize\itshape, anchor=north] at (5,-14) {player 2};

  \foreach \xx/\first/\last in {1/6.2/0, 2/46.9/23.1, 3/65.4/51.5, 4/55.7/56.0, 5/94.7/100} {
    \draw[thick] (\xx-0.07,\first) -- (\xx+0.07,\last);
    \draw[black, fill=white] (\xx-0.07,\first) circle (2.4pt);
    \fill[black] (\xx+0.07,\last) circle (2.4pt);
  }

  \node[font=\scriptsize, anchor=south, text=cursedcol] at (1,110) {betting};
  \node[font=\scriptsize, anchor=south, text=cursedcol] at (2,110) {no-betting iff $\chi\geq\tfrac{29}{30}$};
  \draw[thin, cursedcol] (2.7,113) -- (2.7,116) -- (4.3,116) -- (4.3,113);
  \node[font=\scriptsize, anchor=south, text=cursedcol] at (3.5,117) {no-betting iff $\chi\geq\tfrac{7}{8}$};
  \node[font=\scriptsize, anchor=south, text=cursedcol] at (5,110) {no-betting};
  \node[font=\scriptsize\itshape, anchor=south, text=cursedcol] at (3,131) {cautious $\chi$-cursed rationalizability rules out:};

  \begin{scope}[shift={(0.6,-30)}, x=1cm, y=1cm]
    \draw[black, fill=white] (0,0) circle (2.1pt);
    \node[anchor=west, font=\scriptsize] at (0.15,0) {rounds 1--12};
    \fill[black] (2.4,0) circle (2.1pt);
    \node[anchor=west, font=\scriptsize] at (2.55,0) {rounds 13--24};
  \end{scope}
\end{tikzpicture}
\end{center}
\caption{Betting rates by type, from \citet[Table~3]{sovik2009}. Types are ordered by the degree of cursedness at which no-betting is ruled out, as in Proposition~\ref{prop:sovikexp}(vi).}
\label{fig:sovik}
\end{figure}
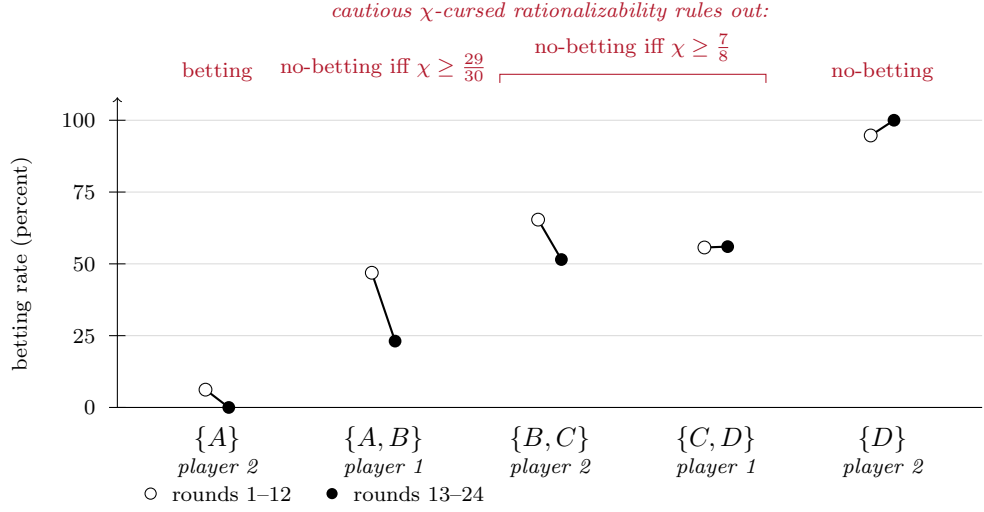

These predictions shed light on the betting rates in the data, which we reproduce in Figure~\ref{fig:sovik}. A substantial share of types $\{A,B\}$, $\{B,C\}$, and $\{C,D\}$ bet, though at $\{A,B\}$ the rate halves in later rounds. Figure~\ref{fig:sovik} orders the types by the degree of cursedness at which no-betting is ruled out, as in clause (vi) of Proposition~\ref{prop:sovikexp}. The lower the theoretical threshold $\chi$, the higher the empirical betting rate.

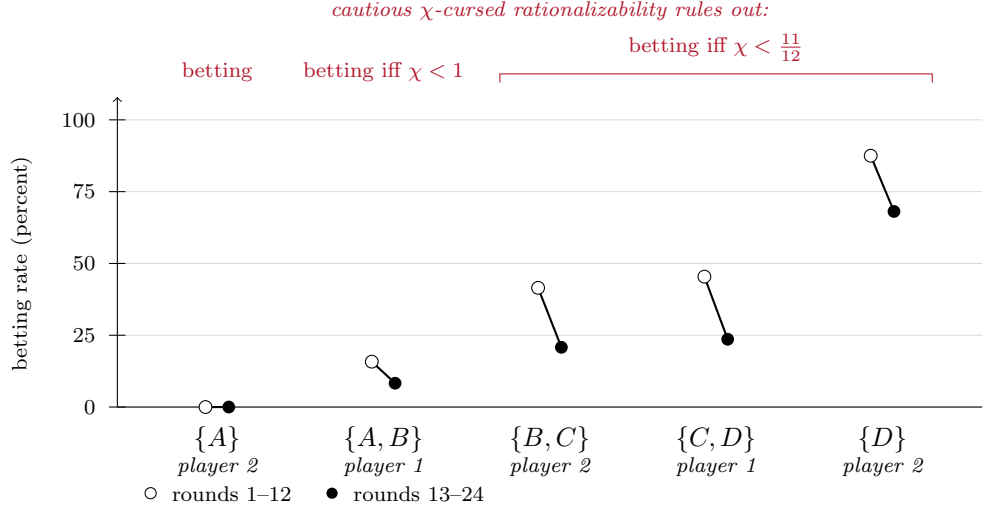
\begin{figure}[H]
\begin{center}
\begin{tikzpicture}[x=2.2cm, y=0.038cm]
  \draw[->] (0.4,0) -- (0.4,108);
  \foreach \yy in {0,25,50,75,100} {
    \draw (0.35,\yy) -- (0.45,\yy);
    \node[font=\scriptsize, anchor=east] at (0.33,\yy) {\yy};
    \draw[gray!30, very thin] (0.45,\yy) -- (5.6,\yy);
  }
  \node[font=\scriptsize, rotate=90, anchor=south] at (-0.05,50) {betting rate (percent)};
  \draw (0.4,0) -- (5.6,0);

  \foreach \xx/\lab in {1/{$\{A\}$}, 2/{$\{A,B\}$}, 3/{$\{B,C\}$}, 4/{$\{C,D\}$}, 5/{$\{D\}$}} {
    \node[font=\small, anchor=north] at (\xx,-3) {\lab};
  }
  \foreach \xx/\pl in {1/2, 2/1, 3/2, 4/1, 5/2} {
    \node[font=\scriptsize\itshape, anchor=north] at (\xx,-14) {player \pl};
  }

  \foreach \xx/\first/\last in {1/0/0, 2/15.8/8.3, 3/41.5/20.8, 4/45.4/23.6, 5/87.5/68.1} {
    \draw[thick] (\xx-0.07,\first) -- (\xx+0.07,\last);
    \draw[black, fill=white] (\xx-0.07,\first) circle (2.4pt);
    \fill[black] (\xx+0.07,\last) circle (2.4pt);
  }

  \node[font=\scriptsize, anchor=south, text=cursedcol] at (1,110) {betting};
  \node[font=\scriptsize, anchor=south, text=cursedcol] at (2,110) {betting iff $\chi<1$};
  \draw[thin, cursedcol] (2.7,113) -- (2.7,116) -- (5.3,116) -- (5.3,113);
  \node[font=\scriptsize, anchor=south, text=cursedcol] at (4,117) {betting iff $\chi<\tfrac{11}{12}$};
  \node[font=\scriptsize\itshape, anchor=south, text=cursedcol] at (3,131) {cautious $\chi$-cursed rationalizability rules out:};

  \begin{scope}[shift={(0.6,-30)}, x=1cm, y=1cm]
    \draw[black, fill=white] (0,0) circle (2.1pt);
    \node[anchor=west, font=\scriptsize] at (0.15,0) {rounds 1--12};
    \fill[black] (2.4,0) circle (2.1pt);
    \node[anchor=west, font=\scriptsize] at (2.55,0) {rounds 13--24};
  \end{scope}
\end{tikzpicture}
\end{center}
\caption{Betting rates by type in the strict-dominance design, from \citet[Table~3]{sovik2009}. Types are ordered as in Figure~\ref{fig:sovik}.}
\label{fig:sovikstrict}
\end{figure}

\citet{sovik2009} has an alternative treatment (the \emph{strict-dominance design}), in which a player who chooses to bet but whose counterparty declines is paid $0$ instead of $1$. This change makes the game solvable by iterated strict dominance, and classical rationalizability predicts that every type does not bet. Empirical betting rates, displayed by type in Figure~\ref{fig:sovikstrict}, are lower in this treatment over all 24 rounds, with types $\{A,B\}$, $\{B,C\}$, and $\{C,D\}$ reducing their betting by 20 to 27 percentage points.

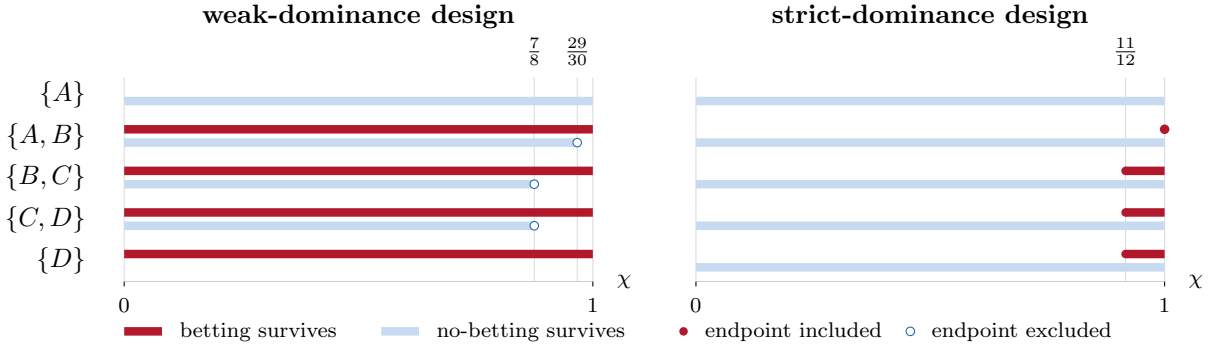
\begin{figure}[H]
\begin{center}
\begin{tikzpicture}[x=6.2cm, y=0.55cm]
  \def\ybar{0.16}
  \def\barw{3.2pt}
  \foreach \row/\lab in {4/{$\{A\}$}, 3/{$\{A,B\}$}, 2/{$\{B,C\}$}, 1/{$\{C,D\}$}, 0/{$\{D\}$}} {
    \node[font=\small, anchor=east] at (-0.06,\row) {\lab};
  }
  \begin{scope}
    \node[font=\small\bfseries, anchor=south] at (0.5,5.35) {weak-dominance design};
    \draw[gray!40, very thin] (0,-0.5) -- (1,-0.5);
    \foreach \t/\l in {0/{$0$}, 1/{$1$}} {
      \draw (\t,-0.5) -- (\t,-0.62); \node[font=\scriptsize, anchor=north] at (\t,-0.66) {\l};
      \draw[gray!30, very thin] (\t,-0.5) -- (\t,4.4);
    }
    \foreach \t/\l in {0.875/{$\tfrac{7}{8}$}, 0.9667/{$\tfrac{29}{30}$}} {
      \draw[gray!30, very thin] (\t,-0.5) -- (\t,4.4);
      \node[font=\scriptsize, anchor=south] at (\t,4.45) {\l};
    }
    \node[font=\scriptsize, anchor=west] at (1.03,-0.5) {$\chi$};
    \draw[classicalbar, line width=\barw] (0,4-\ybar) -- (1,4-\ybar);
    \draw[cursedcol, line width=\barw] (0,3+\ybar) -- (1,3+\ybar);
    \draw[classicalbar, line width=\barw] (0,3-\ybar) -- (0.9667,3-\ybar);
    \draw[classicalcol, fill=white, thin] (0.9667,3-\ybar) circle (1.6pt);
    \foreach \r in {2,1} {
      \draw[cursedcol, line width=\barw] (0,\r+\ybar) -- (1,\r+\ybar);
      \draw[classicalbar, line width=\barw] (0,\r-\ybar) -- (0.875,\r-\ybar);
      \draw[classicalcol, fill=white, thin] (0.875,\r-\ybar) circle (1.6pt);
    }
    \draw[cursedcol, line width=\barw] (0,0+\ybar) -- (1,0+\ybar);
  \end{scope}
  \begin{scope}[shift={(1.22,0)}]
    \node[font=\small\bfseries, anchor=south] at (0.5,5.35) {strict-dominance design};
    \draw[gray!40, very thin] (0,-0.5) -- (1,-0.5);
    \foreach \t/\l in {0/{$0$}, 1/{$1$}} {
      \draw (\t,-0.5) -- (\t,-0.62); \node[font=\scriptsize, anchor=north] at (\t,-0.66) {\l};
      \draw[gray!30, very thin] (\t,-0.5) -- (\t,4.4);
    }
    \foreach \t/\l in {0.9167/{$\tfrac{11}{12}$}} {
      \draw[gray!30, very thin] (\t,-0.5) -- (\t,4.4);
      \node[font=\scriptsize, anchor=south] at (\t,4.45) {\l};
    }
    \node[font=\scriptsize, anchor=west] at (1.03,-0.5) {$\chi$};
    \draw[classicalbar, line width=\barw] (0,4-\ybar) -- (1,4-\ybar);
    \draw[classicalbar, line width=\barw] (0,3-\ybar) -- (1,3-\ybar);
    \fill[cursedcol] (1,3+\ybar) circle (1.8pt);
    \foreach \r in {2,1,0} {
      \draw[classicalbar, line width=\barw] (0,\r-\ybar) -- (1,\r-\ybar);
      \draw[cursedcol, line width=\barw] (0.9167,\r+\ybar) -- (1,\r+\ybar);
      \fill[cursedcol] (0.9167,\r+\ybar) circle (1.6pt);
    }
  \end{scope}
  \begin{scope}[shift={(0,-1.7)}, x=1cm, y=1cm]
    \draw[cursedcol, line width=\barw] (0,0) -- (0.5,0);
    \node[anchor=west, font=\scriptsize] at (0.6,0) {betting survives};
    \draw[classicalbar, line width=\barw] (3.4,0) -- (3.9,0);
    \node[anchor=west, font=\scriptsize] at (4.0,0) {no-betting survives};
    \fill[cursedcol] (7.4,0) circle (1.6pt);
    \node[anchor=west, font=\scriptsize] at (7.55,0) {endpoint included};
    \draw[classicalcol, fill=white, thin] (10.4,0) circle (1.6pt);
    \node[anchor=west, font=\scriptsize] at (10.55,0) {endpoint excluded};
  \end{scope}
\end{tikzpicture}
\end{center}
\caption{Values of $\chi$ at which each action survives cautious $\chi$-cursed rationalizability, by design and type. Filled and open dots mark included and excluded endpoints.}
\label{fig:sovikchi}
\end{figure}

The alternative treatment substantially changes the predictions of cautious $\chi$-cursed rationalizability. We report the ranges of $\chi$ for which each action survives, by treatment and type, in Figure~\ref{fig:sovikchi}. In the weak-dominance design, betting survives at all degrees of cursedness, for all types except $\{A\}$. By contrast, in the strict-dominance design, betting survives only for highly cursed players. Thus, a population of players with heterogeneous $\chi$ appears broadly consistent with the treatment effect reported by \citet{sovik2009}.

\section{Conclusion}

We close by discussing some limitations of cursed rationalizability and issues of interpretation.

Cursed response captures the strategic manifestation of a broader cognitive error. Similar errors arise even in one-player decision problems \citep{charnesslevin2009} and when human players play against their own past behavior \citep{nagel2025}. Cursed rationalizability does not account for these non-strategic anomalies, and it seems difficult to capture them in a parsimonious model grounded in game-theoretic primitives.

As with other cognitive errors, people's propensity for mistakes depends on attention and context. For instance, it is plausible that players are less cursed when reasoning about observed events than about hypotheticals \citep{cohenli2026}, and that their behavior will respond to changes in framing \citep{espondavespa2024} and stakes \citep{alaouipenta2016}, in ways that cursed rationalizability does not capture.

Our preferred interpretation of cursed response is not that players hold mistaken beliefs, but rather that they neglect the link between their own payoffs and others players' information, behaving as if they are in the virtual game instead of the actual game. Under cursed equilibrium, both interpretations are tenable. But cursed rationalizability is not an equilibrium theory: there is no distribution of play against which a conjecture could be called mistaken.

When faced with a complex strategic situation, a cursed player simplifies by reasoning about one part at a time. For example, in an auction, she considers how her own evaluation affects her bid, and how her opponent's evaluation affects their bid, but neglects the link from her opponent's evaluation to her own. 

Without formal training, bidders may not grasp the subtle distinction between private and common values. And even economists abstract away from information linkages \citep{myerson1981}, although few real-world settings have purely private values. Cursed response amounts to the hypothesis that bidders also neglect those linkages, though perhaps more often and less fruitfully than we do. As we have shown, this kind of neglect is separate from the hypothesis of equilibrium.

\appendix

\section{Omitted proofs}

\label{app:proofs}

\subsection{Proof of Proposition~\ref{prop:pennies}}\label{app:proof-pennies}

\begin{proof}
Three observations first. (a) Row is informed: her posterior over $(\theta, t_2)$ is degenerate, so every conjecture of hers equals its cursed projection and her evaluation is classical at every $\chi$; calling against her type pays $-8 < 1$ against everything and is gone after stage one. (b) The state is measurable with respect to the conditioning information on both sides---Row observes it, and Column's posterior concentrates on $t_1 = \theta$---so conditional on the opponent's type there is nothing left for actions to correlate with: every conjecture is partially correlated, the two recursions coincide stage by stage at every $\chi$, and (3) holds. A conjecture for Column is thus a pair $(\sigma_H, \sigma_T)$ of conjectured type mixtures, with action marginal $\nu = \tfrac12\sigma_H + \tfrac12\sigma_T$. (c) By the evaluation identity of Proposition~\ref{prop:chiresp}, Column's $\chi$-cursed value of a call is $(1-\chi)$ times its true expected payoff plus $\chi$ times its virtual payoff against $\nu$.

\emph{Stage one.} Type $H$ retains exactly $\{h, o\}$: $h$ is the strict best response to $\delta_h$ (worth $4 > 1$), $o$ to $\delta_t$, and $t$ is dominated; type $T$ symmetrically retains $\{t, o\}$. Column retains everything at every $\chi$: against the conjecture that both types call $t$, his call $h$ is worth $4$ at every $\chi$ (true value $4$, virtual value $4$); symmetrically for his call $t$; and $o$ is a best response to the truthful conjecture---each type calling its state---under which both of his calls are worth $0$ at every $\chi$.

\emph{Stage two.} Licensing now pairs type $H$ only with $\{h, o\}$ and type $T$ only with $\{t, o\}$. Under every licensed conjecture the true expected payoff of each of Column's calls is \emph{zero}: a call pays nonzero only in the false-call cells $(\theta, a_1) = (H, t)$ and $(T, h)$, and the false calls were removed in stage one, so both cells carry probability zero. So the $\chi$-cursed value of the call $h$ is $\chi$ times its virtual value: writing $\alpha := \sigma_H(h)$ and $\beta := \sigma_T(t)$, the marginal is $\nu(h) = \tfrac{\alpha}{2}$, $\nu(t) = \tfrac{\beta}{2}$, and the call is worth $2\chi(\beta - \alpha)$, maximized at $2\chi$ by the witness $(\alpha, \beta) = (0, 1)$: type $T$ calls, type $H$ stays out. The call $t$ is symmetric, and $o$ pays $1$. If $\chi < \tfrac12$, both calls fall short of $1$ against every licensed conjecture and Column's stage-two set is $\{o\}$. If $\chi \geq \tfrac12$, both calls survive as $\chi$-cursed responses to their witnesses---weak ones at the threshold $\chi = \tfrac12$, which the recursion retains. Row's sets are unchanged at stage two.

\emph{Convergence.} For $\chi < \tfrac12$: with Column held to $\{o\}$, a true call pays $0 < 1$, so stage three deletes it for both types, and the recursion is constant thereafter. For $\chi \geq \tfrac12$: the stage-one family supports itself---the true calls best respond to $\delta_h$ and $\delta_t$, each type's $o$ to $\delta_o$, Column's calls to their witnesses, his $o$ to the truthful conjecture---so it has the $\chi$-cursed response property and survives in full by Corollary~\ref{cor:chifound}(iv); since it already equals the stage-one sets, the recursion is constant from stage one on. This proves (2).

\emph{Equilibrium.} Let $\sigma^*$ be a $\chi$-cursed equilibrium and let $x_H := \sigma^*_1(h \mid H)$ and $x_T := \sigma^*_1(t \mid T)$; by (a), the rest of each type's probability sits on $o$. Column's true conjecture assigns probability zero to false calls, so, as at stage two, his equilibrium values are $2\chi(x_T - x_H)$ for the call $h$, $2\chi(x_H - x_T)$ for the call $t$, and $1$ for $o$. If $x_H = x_T$: both calls are worth $0 < 1$, so $\sigma^*_2 = \delta_o$ strictly; a true call then pays $0 < 1$, forcing $x_H = x_T = 0$---the all-$o$ profile, which is indeed an equilibrium. If $x_H > x_T$: the call $h$ is worth at most $0 < 1$, so $\sigma^*_2 \in \Delta(\{t, o\})$, say with weight $q$ on $t$; type $H$'s true call then pays $-4q \leq 0 < 1$, so $x_H = 0$, contradicting $x_H > x_T \geq 0$. The case $x_T > x_H$ is symmetric. This proves (1), mixing included, at every $\chi$.
\end{proof}

\subsection{Proof of Proposition~\ref{prop:akexp}}\label{app:proof-akexp}

\begin{proof}
Fix a bidder $i$ with signal $x$, let $j$ denote the opponent, and let $n:=|S|$. Winning at price $a$ against opponent signal $y$ pays $x+y+K_{i}-a$; write $\mathrm{Win}(b,a)$ for the probability that bid $b$ wins against bid $a$ ($1$ if $b>a$, $\tfrac{1}{2}$ if $b=a$, $0$ otherwise). A conjecture assigns to each opponent signal $y$ a distribution $\sigma(\cdot\mid y)$ over $\mathcal{B}$, and bid $b$ then earns $\tfrac{1}{n}\sum_{y}\sum_{a}\sigma(a\mid y)\,(x+y+K_{i}-a)\,\mathrm{Win}(b,a)$. Because $S$ is symmetric about $2.5$, $B_{\mathrm{EV}}(x)=x+2.5+K_{i}$, which is the value at $y=2.5$ and hence lies in $\mathcal{B}$; so do $\underline{m}_{i}(x)$ and $\overline{m}_{i}(x)$. Since $\mathcal{B}$ contains the values of both bidders, $\min \mathcal{B}\leq 2$ and $\max \mathcal{B}\geq 8$.

(i) Two conjectures suffice. If every opponent signal bids $\max \mathcal{B}$, then every bid below $\max \mathcal{B}$ loses for sure and earns $0$, while bidding $\max \mathcal{B}$ ties at a price that is at least every value and strictly above the average value $x+2.5+K_{i}\leq 7.5$, and so earns a negative amount; every bid below $\max \mathcal{B}$ is therefore a best response. If every opponent signal bids $\min \mathcal{B}$, then every bid above $\min \mathcal{B}$ wins at price $\min \mathcal{B}\leq 2$ and earns $x+2.5+K_{i}-\min \mathcal{B}>0$, the same amount for all such bids; so $\max \mathcal{B}$ is a best response. Both conjectures are degenerate in the opponent's bid, hence coincide with their cursed projections, and the same bids are cursed responses. Nothing is deleted in the first round, so nothing is deleted in any round.

(ii) Under the cursed projection, a conjecture enters bidder $i$'s evaluation only through its bid marginal $\rho$, and bid $b$ is evaluated at $\sum_{a}\rho(a)\,(B_{\mathrm{EV}}(x)-a)\,\mathrm{Win}(b,a)$: a private-value second-price auction with value $B_{\mathrm{EV}}(x)\in \mathcal{B}$. For $b\neq B_{\mathrm{EV}}(x)$, bidding $B_{\mathrm{EV}}(x)$ weakly dominates $b$, with a strict gain when the opponent bids exactly $b$: if $b<B_{\mathrm{EV}}(x)$, a tie at a profitable price is won outright; if $b>B_{\mathrm{EV}}(x)$, half of a loss is avoided. A cautious conjecture has full support on $\mathcal{B}$ for every opponent signal, so $\rho(b)>0$ and $b$ is not a cursed response. $B_{\mathrm{EV}}(x)$ is a cursed response to every conjecture, and remains one when the opponent's licensed sets are the singletons $\{B_{\mathrm{EV}}(y)\}$. So the cautious round leaves exactly $B_{\mathrm{EV}}(x)$, and the hierarchy is constant thereafter.

(iii) \emph{Bids outside the value range are deleted in the cautious round.} Let $b>\overline{m}_{i}(x)$. Against an opponent bid $a$, replacing $b$ by $\overline{m}_{i}(x)$ changes the payoff only for $a\in[\overline{m}_{i}(x),b]$, where $b$ wins or ties at a price at least the largest possible value: the change is nonnegative for every $(y,a)$ and strictly positive at $a=b$ for every $y$. A full-support conjecture puts positive weight on the opponent bidding $b$, so $b$ earns strictly less than $\overline{m}_{i}(x)$ and is not a best response. Symmetrically, for $b<\underline{m}_{i}(x)$, replacing $b$ by $\underline{m}_{i}(x)$ gains for $a\in[b,\underline{m}_{i}(x))$---a win at a price below the smallest possible value---strictly at $a=b$.

\emph{Bids inside the value range survive the cautious round.} Fix $b\in[\underline{m}_{i}(x),\overline{m}_{i}(x)]\cap \mathcal{B}$. Suppose first that $\min \mathcal{B}<b<\max \mathcal{B}$, and let $a_{1}$ and $a_{2}$ be the bids in $\mathcal{B}$ adjacent to $b$ from below and from above. Consider the conjecture $\nu$ in which opponent signal $4$ bids $a_{1}$, opponent signal $1$ bids $a_{2}$, and every other opponent signal $y$ bids its break-even price $x+y+K_{i}\in \mathcal{B}$. Under $\nu$, every bid earns zero against the break-even bidders. Against signal $4$, $b$ beats $a_{1}$ outright and earns $\tfrac{1}{n}(\overline{m}_{i}(x)-a_{1})>0$; a lower bid either ties at $a_{1}$ and earns half of this, or loses and earns nothing. Against signal $1$, $b$ loses to $a_{2}$ and earns exactly zero, whereas any higher bid wins or ties at the price $a_{2}$, which exceeds the value $\underline{m}_{i}(x)$ against that signal, and so loses at least $\tfrac{1}{2n}(a_{2}-\underline{m}_{i}(x))>0$. Hence $b$ beats every other bid by some $\gamma>0$ under $\nu$. Let $\eta$ give every opponent signal the uniform distribution on $\mathcal{B}$, and let $\mu:=(1-\kappa)\nu+\kappa\eta$. Then $\mu$ has full support on $\mathcal{B}$ for every opponent signal, payoffs are linear in the conjecture, and for $\kappa$ small enough $b$ is the unique best response to $\mu$. So $b$ survives the cautious round. It remains to treat $b=\min \mathcal{B}$ and $b=\max \mathcal{B}$. If $b=\min \mathcal{B}$ there is no lower bid to deter; let signal $4$ bid its break-even price $\overline{m}_{i}(x)$ instead of $a_{1}$, so that it too is payoff-neutral, and the argument for higher bids applies unchanged. Symmetrically, if $b=\max \mathcal{B}$, let signal $1$ bid $\underline{m}_{i}(x)$ instead of $a_{2}$.

\emph{No later round deletes anything.} The witness $\mu$ above puts weight on bids outside the opponent's surviving sets, so later rounds need a different one. After the cautious round, the licensed set for opponent signal $y$ is exactly $[\underline{m}_{j}(y),\overline{m}_{j}(y)]\cap \mathcal{B}=[y+1+K_{j},\,y+4+K_{j}]\cap \mathcal{B}$, and we show that each $b\in[\underline{m}_{i}(x),\overline{m}_{i}(x)]\cap \mathcal{B}$ is a best response to some conjecture on these licensed sets. If $1+K_{j}\leq x+K_{i}\leq 4+K_{j}$ (always in the standard treatment; in the private-value-advantage treatment, for the advantaged bidder with $x\leq 3$ and for the disadvantaged bidder with $x\geq 2$), the break-even bid $x+y+K_{i}$ of every opponent signal $y$ lies in its licensed set, and against the conjecture in which each signal bids it every bid earns zero, so $b$ is a best response. If $x+K_{i}>4+K_{j}$, every price in the licensed sets is below bidder $i$'s value against every opponent signal, so winning is always profitable; against the conjecture in which each signal $y$ bids the minimum $y+1+K_{j}$ of its licensed set, the bid $b\geq\underline{m}_{i}(x)>5+K_{j}$ wins every auction, as does every higher bid, while every lower bid forgoes some profitable win; so $b$ is a best response. If $x+K_{i}<1+K_{j}$, every price in the licensed sets exceeds bidder $i$'s value against every opponent signal, so winning is always a loss; against the conjecture in which each signal $y$ bids the maximum $y+4+K_{j}$ of its licensed set, the bid $b\leq\overline{m}_{i}(x)<5+K_{j}$ loses every auction and earns zero, the most any bid can earn; so $b$ is a best response.
\end{proof}

\clearpage
\section{Exact predictions in the Holt--Sherman designs}

\label{app:hsexp}

Table~\ref{tab:hsexp} reports the values plotted in Figure~\ref{fig:hsexp}. The rationalizability rows apply the cautious refinement; in the equilibrium rows, the seller accepts at her threshold value.

\begin{table}[H]
\begin{center}
{\small
\begin{tabular}{@{}lccc@{}}
\toprule
 & Winner's curse & Loser's curse & No curse \\
\midrule
Mean bids (empirical) & $3.78$ & $0.74$ & $2.03$ \\
Fully cursed equilibrium & $3.56$ & $0.81$ & $\{1.99,2.00\}$ \\
Cursed rationalizability & $[3.42,3.70]$ & $[0.76,0.86]$ & $[1.90,2.09]$ \\
Bayesian Nash equilibrium & $\{2.99,3.00\}$ & $\{0.99,1.00\}$ & $\{1.99,2.00\}$ \\
Interim correlated rationalizability & $[2.76,3.23]$ & $[0.86,1.00]$ & $[1.80,2.19]$ \\
\bottomrule
\end{tabular}
}
\end{center}
\caption{Mean bids from \citet[Table~2]{holtsherman1994} and the predictions of each solution concept on the discretized design.}
\label{tab:hsexp}
\end{table}

\section{Cursed rationalizability in bilateral trade}

\label{app:egr}

\citet{eystergagnonbartschrabin2026} study bilateral trade between two cursed traders, $A$ and $B$. The traders hold possibly different full-support priors $\mu_A$ and $\mu_B$ over a finite state space $E$.\footnote{As noted in Section~\ref{sec:concept}, our definitions extend to type-specific posteriors, so the comparison below does not require a common prior.} At state $e$, trader $i\in\{A,B\}$ learns the cell $E_i(e)$ of her information partition. A \textbf{trade} is a function $t:E\to\mathbb{R}$ specifying the transfer from $B$ to $A$. Each trader chooses Accept or Reject. If both accept in state $e$, then $A$ receives $t(e)$ and $B$ receives $-t(e)$; otherwise both receive zero. In the notation of Section~\ref{sec:concept}, the state is $e$ and the types are the information cells. Write $n_{A}(e):=\mathbb{E}_{\mu_A}[t\mid E_{A}(e)]$ and $n_{B}(e):=\mathbb{E}_{\mu_B}[-t\mid E_{B}(e)]$ for the traders' \emph{na\"{\i}ve} expected gains from trade, conditional on their own information alone. Their Definition~A.1 calls the trade $t$ \textbf{cursed rationalizable in state $e$} if $n_{A}(e)>0$ and $n_{B}(e)>0$: each trader strictly prefers to trade conditional on the other's agreement, without drawing inferences about the state from that agreement.

Their criterion applies to a trade at a state, whereas our definition applies to each type's actions. To compare the two, we ask whether Accept survives for both traders' types at that state. Two issues distinguish the concepts. First, our unrefined hierarchy can retain Accept even when the trader's na\"{\i}ve expected gain is negative, because certainty that the opponent rejects makes both actions equally valuable. Second, their criterion requires strict profitability, whereas our best-response condition permits indifference. Proposition~\ref{prop:egr} shows that, once caution rules out the first justification and nonzero expected gains remove the second issue, the two concepts identify the same states for trade. Although their criterion involves no iteration, one cautious round already determines the entire hierarchy in this environment.

\begin{example}[A distinguishing trade]\label{ex:egrdistinct} There are two equally likely states. Trader $A$ observes the state and $B$ does not; $t(1)=1$ and $t(2)=-3$. Then $n_{A}(1)=1$, $n_{A}(2)=-3$, and $n_{B}(1)=n_{B}(2)=1$, so in their sense the trade is cursed rationalizable in state $1$ only. Under Definition~\ref{def:typewise}, however, both actions survive for every type at every round: against the conjecture that the opponent rejects, Accept and Reject both pay zero, so both are cursed responses. In particular, $A$ acceptance in state~$2$ survives despite her negative expected gain. The profile in which every type accepts therefore consists entirely of cursed-rationalizable actions and generates trade in both states.\end{example}

The example illustrates the veto problem of Section~\ref{sec:experiments}. Our cautious refinement addresses it by requiring first-round conjectures to assign positive probability to every opponent action.

\begin{proposition}\label{prop:egr} Consider a bilateral trading game in which $n_{A}(e)\neq0$ and $n_{B}(e)\neq0$ for every state $e$. The first cautious round selects a unique action for every type: trader $i$ accepts if $n_i(e)>0$ and rejects if $n_i(e)<0$. No subsequent round eliminates any further action. Hence trade occurs in state $e$ under cautious cursed rationalizability if and only if $t$ is cursed rationalizable in state $e$ in the sense of \citet{eystergagnonbartschrabin2026}. \end{proposition}

\begin{proof} Fix $A$'s type $E_{A}(e)$ and a conjecture $\mu$, and let $\rho$ be the probability $\mu$ assigns to $B$ accepting. Under the cursed projection, $B$'s action is paired with the state independently, so $A$'s cursed value of Accept is $\rho\,n_{A}(e)$ and that of Reject is $0$. In the cautious round $\mu$ has full support on $B$'s actions, so $\rho\in(0,1)$, and Accept is the unique cursed response if $n_{A}(e)>0$, Reject if $n_{A}(e)<0$. In later rounds the licensed sets are these singletons and $\rho\in[0,1]$; the selected action remains a cursed response ($\rho\,n_{A}(e)\geq0$ in the first case, $\leq0$ in the second), so nothing changes. The same holds for $B$ with $n_{B}(e)$. Trade occurs in state $e$ if and only if both types accept, that is, if and only if $n_{A}(e)>0$ and $n_{B}(e)>0$. \end{proof}

The genericity condition excludes exact indifference, where the two conventions part: their definition uses strict inequalities, while the cautious recursion retains both actions for an indifferent type. \citeauthor{eystergagnonbartschrabin2026} also observe that $t$ is cursed rationalizable in state $e$ if and only if both traders accept $t$ in state $e$ in some fully cursed equilibrium. By Proposition~\ref{prop:egr}, the equilibrium hypothesis is generically not needed: one cautious round delivers the same trades, as in the betting game of Section~\ref{sec:sovikexp}.


\begin{thebibliography}{Crawford et~al.(2013)Crawford, Costa-Gomes, and Iriberri}
\bibitem[Alaoui and Penta(2016)]{alaouipenta2016} Alaoui, L. and Penta, A. (2016). \newblock Endogenous depth of reasoning. \newblock \textit{Review of Economic Studies}, 83(4):1297--1333.

\bibitem[Avery and Kagel(1997)]{averykagel1997} Avery, C. and Kagel, J. H. (1997). \newblock Second-price auctions with asymmetric payoffs: An experimental investigation. \newblock \textit{Journal of Economics \& Management Strategy}, 6(3):573--603.

\bibitem[Ball et~al.(1991)Ball, Bazerman, and Carroll]{ballbazermancarroll1991} Ball, S.~B., Bazerman, M.~H., and Carroll, J.~S. (1991). \newblock An evaluation of learning in the bilateral winner's curse. \newblock \textit{Organizational Behavior and Human Decision Processes}, 48(1):1--22.

\bibitem[Bernheim(1984)]{bernheim1984} Bernheim, B.~D. (1984). \newblock Rationalizable strategic behavior. \newblock \textit{Econometrica}, 52(4):1007--1028.

\bibitem[Bikhchandani(1988)]{bikhchandani1988} Bikhchandani, S. (1988). \newblock Reputation in repeated second-price auctions. \newblock \textit{Journal of Economic Theory}, 46(1):97--119.

\bibitem[Camerer et~al.(2004)Camerer, Ho, and Chong]{camererhochong2004}
Camerer, C.~F., Ho, T.-H., and Chong, J.-K. (2004).
\newblock A cognitive hierarchy model of games.
\newblock \textit{Quarterly Journal of Economics}, 119(3):861--898.

\bibitem[Charness and Levin(2009)]{charnesslevin2009} Charness, G. and Levin, D. (2009). \newblock The origin of the winner's curse: A laboratory study. \newblock \textit{American Economic Journal: Microeconomics}, 1(1):207--236.

\bibitem[Cohen and Li(2026)]{cohenli2026} Cohen, S. and Li, S. (2026). \newblock Sequential cursed equilibrium. \newblock \textit{American Economic Review}, 116(3):934--976.

\bibitem[Crawford et~al.(2013)Crawford, Costa-Gomes, and Iriberri]{crawfordcostagomesiriberri2013} Crawford, V.~P., Costa-Gomes, M.~A., and Iriberri, N. (2013). \newblock Structural models of nonequilibrium strategic thinking: Theory, evidence, and applications. \newblock \textit{Journal of Economic Literature}, 51(1):5--62.

\bibitem[Crawford and Iriberri(2007)]{crawfordiriberri2007} Crawford, V.~P. and Iriberri, N. (2007). \newblock Level-$k$ auctions: Can a nonequilibrium model of strategic thinking explain the winner's curse and overbidding in private-value auctions? \newblock \textit{Econometrica}, 75(6):1721--1770.

\bibitem[Dekel and Fudenberg(1990)]{dekelfudenberg1990} Dekel, E. and Fudenberg, D. (1990). \newblock Rational behavior with payoff uncertainty. \newblock \textit{Journal of Economic Theory}, 52(2):243--267.

\bibitem[Dekel et~al.(2007)]{dekelfudenbergmorris2007} Dekel, E., Fudenberg, D., and Morris, S. (2007). \newblock Interim correlated rationalizability. \newblock \textit{Theoretical Economics}, 2(1):15--40.

\bibitem[Dyer et~al.(1989)Dyer, Kagel, and Levin]{dyerkagellevin1989} Dyer, D., Kagel, J. H., and Levin, D. (1989). \newblock A comparison of naive and experienced bidders in common value offer auctions: A laboratory analysis. \newblock \textit{The Economic Journal}, 99(394):108--115.

\bibitem[Ely and Peski(2006)]{elypeski2006} Ely, J.~C. and Peski, M. (2006). \newblock Hierarchies of belief and interim rationalizability. \newblock \textit{Theoretical Economics}, 1(1):19--65.

\bibitem[Esponda(2008)]{esponda2008} Esponda, I. (2008). \newblock Behavioral equilibrium in economies with adverse selection. \newblock \textit{American Economic Review}, 98(4):1269--1291.

\bibitem[Esponda and Pouzo(2016)]{espondapouzo2016} Esponda, I. and Pouzo, D. (2016). \newblock Berk--Nash equilibrium: A framework for modeling agents with misspecified models. \newblock \textit{Econometrica}, 84(3):1093--1130.

\bibitem[Esponda and Vespa(2024)]{espondavespa2024} Esponda, I. and Vespa, E. (2024). \newblock Contingent thinking and the sure-thing principle: Revisiting classic anomalies in the laboratory. \newblock \textit{Review of Economic Studies}, 91(5):2806--2831.

\bibitem[Eyster and Rabin(2005)]{eysterrabin2005} Eyster, E. and Rabin, M. (2005). \newblock Cursed equilibrium. \newblock \textit{Econometrica}, 73(5):1623--1672.

\bibitem[Eyster et~al.(2026)Eyster, Gagnon-Bartsch, and Rabin]{eystergagnonbartschrabin2026} Eyster, E., Gagnon-Bartsch, T., and Rabin, M. (2026). \newblock Disagreement, information, and trade. \newblock Working paper, September 4, 2026.

\bibitem[Fong et~al.(2025)]{fonglinpalfrey2025} Fong, M.-J., Lin, P.-H., and Palfrey, T.~R. (2025). \newblock Cursed sequential equilibrium. \newblock \textit{American Economic Review}, 115(8):2616--2658.

\bibitem[Fudenberg(2006)]{fudenberg2006} Fudenberg, D. (2006). \newblock Advancing beyond \textit{Advances in Behavioral Economics}. \newblock \textit{Journal of Economic Literature}, 44(3):694--711.

\bibitem[Holt and Sherman(1994)]{holtsherman1994} Holt, C.~A. and Sherman, R. (1994). \newblock The loser's curse. \newblock \textit{American Economic Review}, 84(3):642--652.

\bibitem[Jehiel(2005)]{jehiel2005} Jehiel, P. (2005). \newblock Analogy-based expectation equilibrium. \newblock \textit{Journal of Economic Theory}, 123(2):81--104.

\bibitem[Jehiel and Koessler(2008)]{jehielkoessler2008} Jehiel, P. and Koessler, F. (2008). \newblock Revisiting games of incomplete information with analogy-based expectations. \newblock \textit{Games and Economic Behavior}, 62(2):533--557.

\bibitem[Kagel(1995)]{kagel1995} Kagel, J.~H. (1995). \newblock Auctions: A survey of experimental research. \newblock In Kagel, J.~H. and Roth, A.~E., editors, \textit{The Handbook of Experimental Economics}, pages 501--585. Princeton University Press, Princeton.

\bibitem[Kagel and Levin(1986)]{kagellevin1986} Kagel, J.~H. and Levin, D. (1986). \newblock The winner's curse and public information in common value auctions. \newblock \textit{American Economic Review}, 76(5):894--920.

\bibitem[Kagel and Levin(2016)]{kagellevin2016} Kagel, J.~H. and Levin, D. (2016). \newblock Auctions: A survey of experimental research. \newblock In Kagel, J.~H. and Roth, A.~E., editors, \textit{The Handbook of Experimental Economics, Volume 2}, pages 563--637. Princeton University Press, Princeton.

\bibitem[Kagel and Richard(2001)]{kagelrichard2001} Kagel, J.~H. and Richard, J.-F. (2001). \newblock Super-experienced bidders in first-price common-value auctions: Rules of thumb, Nash equilibrium bidding, and the winner's curse. \newblock \textit{Review of Economics and Statistics}, 83(3):408--419.

\bibitem[Mart{\'\i}nez-Marquina et~al.(2019)]{martinezmarquinaniederlevespa2019} Mart{\'\i}nez-Marquina, A., Niederle, M., and Vespa, E. (2019). \newblock Failures in contingent reasoning: The role of uncertainty. \newblock \textit{American Economic Review}, 109(10):3437--3474.

\bibitem[Marx and Swinkels(1997)]{marxswinkels1997} Marx, L.~M. and Swinkels, J.~M. (1997). \newblock Order independence for iterated weak dominance. \newblock \textit{Games and Economic Behavior}, 18(2):219--245.

\bibitem[Manili(2024)]{manili2024} Manili, J. (2024). \newblock Order independence for rationalizability. \newblock \textit{Games and Economic Behavior}, 143:152--160.

\bibitem[Milgrom and Roberts(1990)]{milgromroberts1990} Milgrom, P. and Roberts, J. (1990). \newblock Rationalizability, learning, and equilibrium in games with strategic complementarities. \newblock \textit{Econometrica}, 58(6):1255--1277.

\bibitem[Myerson(1981)]{myerson1981} Myerson, R.~B. (1981). \newblock Optimal auction design. \newblock \textit{Mathematics of Operations Research}, 6(1):58--73.

\bibitem[Nagel(1995)]{nagel1995} Nagel, R. (1995). \newblock Unraveling in guessing games: An experimental study. \newblock \textit{American Economic Review}, 85(5):1313--1326.

\bibitem[Nagel et~al.(2025)Nagel, Niederle, and Vespa]{nagel2025}
Nagel, L., Niederle, M., and Vespa, E. (2025).
\newblock Decomposing the winner's curse.
\newblock Working paper, November 10, 2025.

\bibitem[Niederle and Vespa(2023)]{niederlevespa2023} Niederle, M. and Vespa, E. (2023). \newblock Cognitive limitations: Failures of contingent thinking. \newblock \textit{Annual Review of Economics}, 15:307--328.

\bibitem[Pearce(1984)]{pearce1984} Pearce, D.~G. (1984). \newblock Rationalizable strategic behavior and the problem of perfection. \newblock \textit{Econometrica}, 52(4):1029--1050.

\bibitem[Samuelson and Bazerman(1985)]{samuelsonbazerman1985} Samuelson, W. F. and Bazerman, M. H. (1985). \newblock The winner's curse in bilateral negotiations. \newblock \textit{Research in Experimental Economics}, 3:105--138.

\bibitem[Sonsino et~al.(2002)Sonsino, Erev, and Gilat]{sonsinoerevgilat2002} Sonsino, D., Erev, I., and Gilat, S. (2002). \newblock On rationality, learning and zero-sum betting---an experimental study of the no-betting conjecture. \newblock Working paper, Technion.

\bibitem[S{ø}vik(2009)]{sovik2009} S{ø}vik, Y. (2009). \newblock Strength of dominance and depths of reasoning---an experimental study. \newblock \textit{Journal of Economic Behavior \& Organization}, 70(1):196--205.

\bibitem[Spiegler(2016)]{spiegler2016} Spiegler, R. (2016). \newblock Bayesian networks and boundedly rational expectations. \newblock \textit{The Quarterly Journal of Economics}, 131(3):1243--1290.

\bibitem[Spiegler(2020)]{spiegler2020} Spiegler, R. (2020). \newblock Behavioral implications of causal misperceptions. \newblock \textit{Annual Review of Economics}, 12:81--106.

\bibitem[Stahl and Wilson(1994)]{stahlwilson1994} Stahl, D.~O. and Wilson, P.~W. (1994). \newblock Experimental evidence on players' models of other players. \newblock \textit{Journal of Economic Behavior \& Organization}, 25(3):309--327.

\bibitem[Stahl and Wilson(1995)]{stahlwilson1995} Stahl, D.~O. and Wilson, P.~W. (1995). \newblock On players' models of other players: Theory and experimental evidence. \newblock \textit{Games and Economic Behavior}, 10(1):218--254.

\bibitem[Tang(2015)]{tang2015} Tang, Q. (2015). \newblock Interim partially correlated rationalizability. \newblock \textit{Games and Economic Behavior}, 91:36--44.

\bibitem[Van~Zandt(2010)]{vanzandt2010} Van~Zandt, T. (2010). \newblock Interim Bayesian Nash equilibrium on universal type spaces for supermodular games. \newblock \textit{Journal of Economic Theory}, 145(1):249--263.

\end{thebibliography}
\end{document}